# Resistive switching and charge accumulation in $Hf_{0.5}Zr_{0.5}O_2$ nanoparticles

Oleksandr S. Pylypchuk[1], Ihor V. Fesych[2], Victor V. Vainberg[1*], Yuri O. Zagorodniy[3], Victor I. Styopkin[1], Juliya M. Gudenko[1], Irina V. Kondakova[3], Lesya P. Yurchenko[3], Victor N. Pavlikov[3], Anna O. Diachenko[4], Mykhailo M. Koptiev[4], Michail D. Volnyanskii [4], Valentin V. Laguta[3], Eugene A. Eliseev[3†], Mikhail P. Trubitsyn[4,5‡], and Anna N. Morozovska [1§]

[1] Institute of Physics, National Academy of Sciences of Ukraine,46, pr. Nauky, 03028 Kyiv, Ukraine

[2]Taras Shevchenko National University of Kyiv, 01030 Kyiv, Ukraine

[3]Frantsevich Institute for Problems in Materials Science, National Academy of Sciences of Ukraine, 3, str. Omeliana Pritsaka, 03142 Kyiv, Ukraine

[4]Oles Honchar Dnipro National University

[5]Institute of Physics of the Czech Academy of Sciences, Na Slovance 1999/2, 182 00 Prague 8, Czech Republic

**Abstract**

We revealed the resistive switching, negative differential resistance and charge accumulation effects in $Hf_{0.5}Zr_{0.5}O_2$ nanopowders sintered by the auto-combustion sol-gel method and annealed at temperatures from 500°C to 800°C. The fraction of the orthorhombic phase, determined by the X-ray diffraction (XRD), decreases from 91 vol.% to 7 vol.% with an increase in the annealing temperature from 600°C to 800°C. The electron paramagnetic resonance (EPR) spectra reveal the great amount of oxygen vacancies in the annealed samples, at that the decrease of the orthorhombic phase fraction (observed with an increase in the annealing temperature) correlates with a decrease in the intensity of EPR spectral lines associated with the oxygen vacancies and impurities. This indicates the participation of oxygen vacancies and other defects in the formation of the orthorhombic phase in the $Hf_{0.5}Zr_{0.5}O_2$ powders. To explain the results of electrophysical measurements, we compare the features of the current-voltage characteristics with the phase composition of the $Hf_{0.5}Zr_{0.5}O_2$ powders and with the peculiarities of their EPR spectra. The analysis allows us to relate the resistive switching and charge accumulation observed in $Hf_{0.5}Zr_{0.5}O_2$ nanopowders with the appearance of the ferroelectric-like polar regions in the orthorhombic phase of the nanoparticles, which agrees with the calculations performed in the framework of Landau-Ginzburg-Devonshire approach and density functional theory.

* Corresponding author: viktor.vainberg@gmail.com
† Corresponding author: eugene.a.eliseev@gmail.com
‡ Corresponding author: trubitsyn_m@ua.fm
§Corresponding author: anna.n.morozovska@gmail.com

## INTRODUCTION

The discovery of ferroelectric, ferrielectric and antiferroelectric properties in thin films of lead-free hafnium ($HfO_2$) and zirconium ($ZrO_2$) oxides makes them main candidates for the next-generation of Si-compatible ferroelectric memory elements, e.g., the ferroelectric random-access memory (FeRAM) constructed of memristors or/and field-effect transistors (FETs) with ferroelectric gate oxides [1, 2, 3].

Bulk $HfO_2$ and $ZrO_2$ are high-k dielectrics entire the range of available temperatures (up to 1200 K) pressures (up to 15 GPa) [4, 5]. The long-range polar order appears only when the spatial confinement of the $Hf_xZr_{1-x}O_2$ ($0 \leq x \leq 1$) is reduced to nanometer range, when the dipole-dipole interactions and hidden couplings induce the film transition from the non-polar monoclinic phase to the polar orthorhombic phase. Since the orthorhombic phase (space group $Pca2_1$) is metastable, as compared to the bulk monoclinic phase (space group $P2_1/c$), appearance of polar properties in $Hf_xZr_{1-x}O_2$ thin films is highly sensitive to deposition method, substrate/electrode materials, annealing conditions, doping and co-doping [6, 7, 8], and hafnium atoms content “x” [9, 10]. It is generally accepted that the pronounced long-range polar order can exist in a rather limited range of $Hf_xZr_{1-x}O_2$ film thickness, as a rule, from 5 to 50 nm [11, 12]. Recent theoretical [13, 14, 15] and experimental [11, 12, 16] studies revealed a significant influence of the oxygen vacancies, surface and grain boundaries in the appearance of the metastable ferroelectric orthorhombic phase in the $Hf_xZr_{1-x}O_2$ thin films.

The density functional theory (DFT) calculations confirmed that the crystal structure with monoclinic symmetry has the lowest ground state energy in the bulk $Hf_xZr_{1-x}O_2$. At the same time, the DFT calculations [17] revealed that the origin of ferroelectricity in $Hf_xZr_{1-x}O_2$ films is surface-related. Interesting, that two categories of polarization switching paths may exist in $HfO_2$ due to competing monoclinic and orthorhombic phases [18]. As was shown in Ref. [19], the difference between the energies of the polar orthorhombic phases and monoclinic phase in $HfO_2$ structure calculated in the framework of the DFT turned out to be smaller (~20 meV/f.u.) than that calculated by other calculation methods. Thus, the DFT calculations showed that nanoscale $HfO_2$ may become polar, especially in the presence of impurity atoms and/or oxygen vacancies.

In comparison with $Hf_xZr_{1-x}O_2$ thin films, the direct experimental observation of the ferroelectric (or at least ferroelectric-like) properties of $Hf_xZr_{1-x}O_2$ nanoparticles is absent. There exist several observations of the orthorhombic phases (mixture of *Pca*2$_1$, *Pbca* and *Pbcm*) in the small (sizes 3 – 30 nm) $Hf_xZr_{1-x}O_{2-y}$ nanoparticles [19, 20, 21] based on the X-ray diffraction analysis.

However, it was predicted theoretically [22], that ferro-ionic states can emerge in small (less than 20 nm) spherical $Hf_xZr_{1-x}O_2$ nanoparticles covered by ionic-electronic charge, leading to unusual polar, dielectric and electrophysical properties. In this work the phase diagrams, dielectric permittivity, long-range polar and antipolar orderings, and domain structure morphology of the $Hf_xZr_{1-x}O_2$ nanoparticles were calculated using Landau-Ginzburg-Devonshire (LGD) approach. When the concentration of surface charges is relatively low (less than $10^{16}$ cm$^{-2}$), the long-range polar order cannot emerge in $Hf_xZr_{1-x}O_2$ nanoparticles due to the insufficient screening by the ionic-electronic charge. At higher concentrations (~$10^{17}$ – $10^{18}$ cm$^{-2}$) of the surface ions and/or charged vacancies the ferro-ionic coupling supports the long-range polar order in multi-domain $Hf_xZr_{1-x}O_2$ nanoparticles with sizes 5 – 20 nm. The ferro-ionic coupling causes the transition to the single-domain ferro-ionic state at high concentrations (~$10^{19}$ – $10^{20}$ cm$^{-2}$) of the surface charges.

Due to the size effect, the $Hf_xZr_{1-x}O_2$ nanoparticles become paraelectric with an increase in their size above 30 nm, and then dielectric with its further increase above 50 nm [22]. However, it seems reasonable to assume that the ferro-ionic coupling at the surface and/or interfaces can lead to the mixed ionic-electronic conductivity in $Hf_xZr_{1-x}O_{2-y}$ thin films and nanoparticles with the oxygen vacancies (i.e. for y>0), which, as a rule, are present in large amount in these nanomaterials [12, 14, 15]. Emerging reversible dynamics of the oxygen-enriched and oxygen-deficient regions can lead to the resistive switching between metastable states with higher and lower resistance [23]. Like resistive switching in thin films of rare-earth binary oxides $TaO_{2-y}$ [24] and $NbO_{2-y}$ [25], bilayer structure $MgO/MoS_2$ [26], memristive switching [27] is expected in the nanosized $Hf_xZr_{1-x}O_{2-y}$. These nanomaterials may demonstrate negative differential resistance (NDR) region(s) and a pronounced double loop shape of current-voltage curves [28, 29]. In contrast to ferroelectrics and antiferroelectrics, memristors cannot store energy, but they can "remember" the total charge transfer due to the metastable changes of their nonlinear conductance [30]. Resistive switching characteristics are highly sensitive to the electromechanical state of oxide film surfaces [31], which opens a high potential of hafnium-zirconium oxide nanomaterials as controllable working elements of multifunctional nanoelectronic devices and neuromorphic computing.

In this work we revealed experimentally the resistive switching and charge accumulation in $Hf_{0.5}Zr_{0.5}O_2$ nanopowders sintered by the auto-combustion sol-gel method. To explain the experimental results, we analyze phase composition of the nanopowder samples annealed at temperatures from 500°C to 800°C, determined by the X-ray diffraction (XRD) analysis, and relate it with the peculiarities of their electron paramagnetic resonance (EPR) spectra. The analysis allows us to relate the resistive switching and charge accumulation observed in $Hf_{0.5}Zr_{0.5}O_2$ nanopowders with the *possible* appearance of the ferroelectric-like polar regions in the nanoparticles, which agrees with the calculations performed in the framework of LGD approach and DFT.

## EXPERIMENTAL RESULTS

### A. Synthesis of $Hf_{0.5}Zr_{0.5}O_2$ nanopowders and their characterization by X-ray diffraction and scanning electron microscopy

The synthesis of $Hf_{0.5}Zr_{0.5}O_2$ was carried out by the sol-gel method in the presence of the citric acid as a gelling agent. The hafnium and zirconium sources were the crystal hydrates, hafnyl nitrate and zirconyl nitrate, respectively (see **Appendix A** for details). After combustion, the gel was grinded in an agate mortar. The grinded powder was annealed at temperatures (500 – 800)°C. The samples #1 – #4 correspond to $Hf_{0.5}Zr_{0.5}O_2$ nanopowders annealed at temperatures 500°C, 600°C, 700°C and 800°C, respectively. $ZrO_2$ and $HfO_2$ nanopowders, annealed at 800°C, were also studied as reference samples # 5 and #6, respectively. To control the phase composition and study the processes occurring during the synthesis, the method of simultaneous differential thermal analysis and thermogravimetry (DTA/TG) in combination with XRD analysis was used (see **Appendix B** for details).

It appeared that the black color of the sample #1, synthesized at 500°C, evidences the presence of carbon in the composition. This assumption is further confirmed by the X-ray powder diffraction data. Above 620°C, the TG curve reaches a plateau, indicating the formation of an oxide matrix (see **Fig. B1** in **Appendix B**). Other samples have white or light-grey color.

XRD analysis was performed at 290 K on the LabX XRD-6000 diffractometer (Shimadzu, Japan) using CuKα radiation (λ = 1.5406 Å). X-ray diffraction patterns were analyzed by the Rietveld method. The initial expansion basis contains the monoclinic phase (space group P21/c), the tetragonal phase (space group $P4_2$/nmc) and the orthorhombic phases (space groups Pbca, Pbcm and Pca21). As determined from XRD analysis, the $Hf_{0.5}Zr_{0.5}O_2$ samples # 1 – 4 contain the monoclinic phase and/or orthorhombic phase, the $ZrO_2$ sample # 5 contains the monoclinic phase and about 16 vol.% of the tetragonal phase (space group $P4_2$/nmc), the $HfO_2$ sample # 6 contains the 98 % of monoclinic phase and about 2 vol.% of the orthorhombic phase (see **Table B1** in **Appendix B**).The Rietveld-refined diffraction patterns of the zirconium-hafnium oxide powders are shown in **Fig. 1**. The crystallographic parameters, refined by the Rietveld method, are given in **Table B1**.

According to the analysis of X-ray diffractograms, the process of the crystalline phase formation begins at 500°C. However, the hafnium-zirconium oxide still contains carbon inclusions, which are X-ray amorphous; and whose presence worsens the determination of crystallographic parameters and sizes of scattering crystallites. Thus, the temperature of 500°C is insufficient for burning out the remainders of the organic phase in the dry nitrate-citrate gel. In general, the diffractograms of all the studied systems contain diffraction reflections belonging to the orthorhombic and monoclinic phases of $Hf_{0.5}Zr_{0.5}O_2$. The volume fractions of these phases significantly depend on the annealing temperature: the content of the orthorhombic phases fluctuates around 90 vol. % at 500°C and 600°C, and does not exceed 7 vol.% at 800°C. Unfortunately, it appeared impossible to obtain 100 vol. % pure orthorhombic $Hf_{0.5}Zr_{0.5}O_2$.

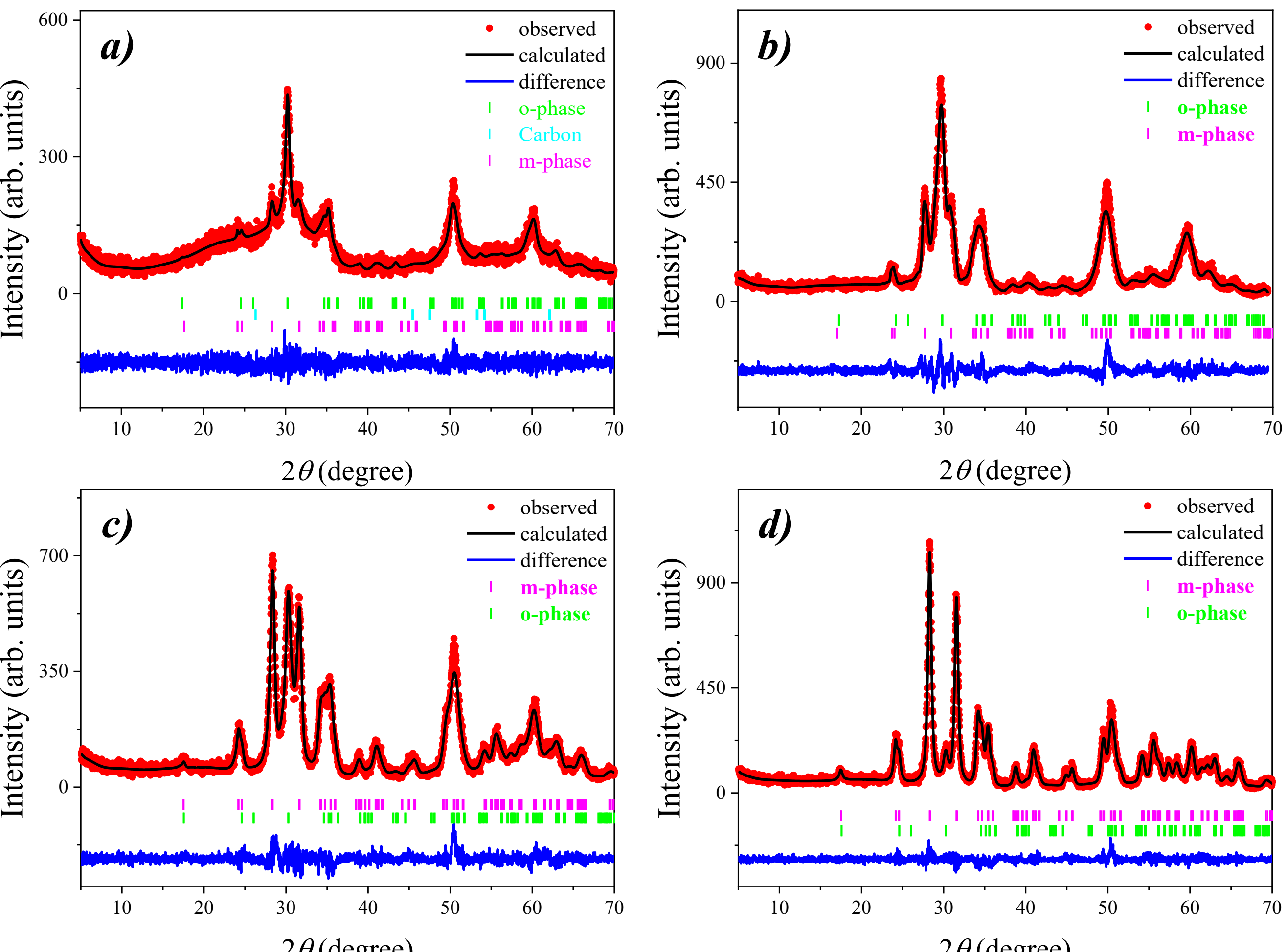

**Figure 1.** XRD pattern of $Hf_{0.5}Zr_{0.5}O_2$ nanopowder samples #1 – 4, synthesized by the sol-gel method at different temperatures: 500°C **(a)**, 600°C **(b)**, 700°C **(c)** and 800°C **(d)**. Experimental data are presented as red circles, black lines correspond to the calculated diffractogram, and their difference is shown in blue color. The vertical green and purple stripes reflect the positions of the Bragg angles of the orthorhombic o-phase and the monoclinic m-phase, respectively.

Scanning electron microscopy (SEM) was performed for a small amount of powder samples placed at a clean silicon wafer (atomic number $Z_{Si}$ = 14). SEM images of the $Hf_{0.5}Zr_{0.5}O_2$ nanopowders, as well as $ZrO_2$ and $HfO_2$ nanopowders, are shown in **Fig. 2.** The brightness of the $Hf_{0.5}Zr_{0.5}O_2$ particles in Z contrast is quite uniform, which indicates the uniformity of their phase composition, i.e., we study a single-phase solid solution. Of course, this statement is true up to the level of 20 − 30 nm resolution inherent to the SEM. Since the resolution of the available SEM set-up cannot be smaller than 30 nm, we cannot observe single nanoparticles of smaller size, however their clusters, as well as submicron particles with sizes less than 200 – 300 nm, can be seen in the images. These results agree with the average size 50 – 100 nm of the coherent scattering regions determined by the XRD analysis. However, the images of materials shown in **Fig. 2**, maybe very different from the shape of individual micro-, meso- and/or nanosized particles sintered by the auto-combustion sol-gel method.

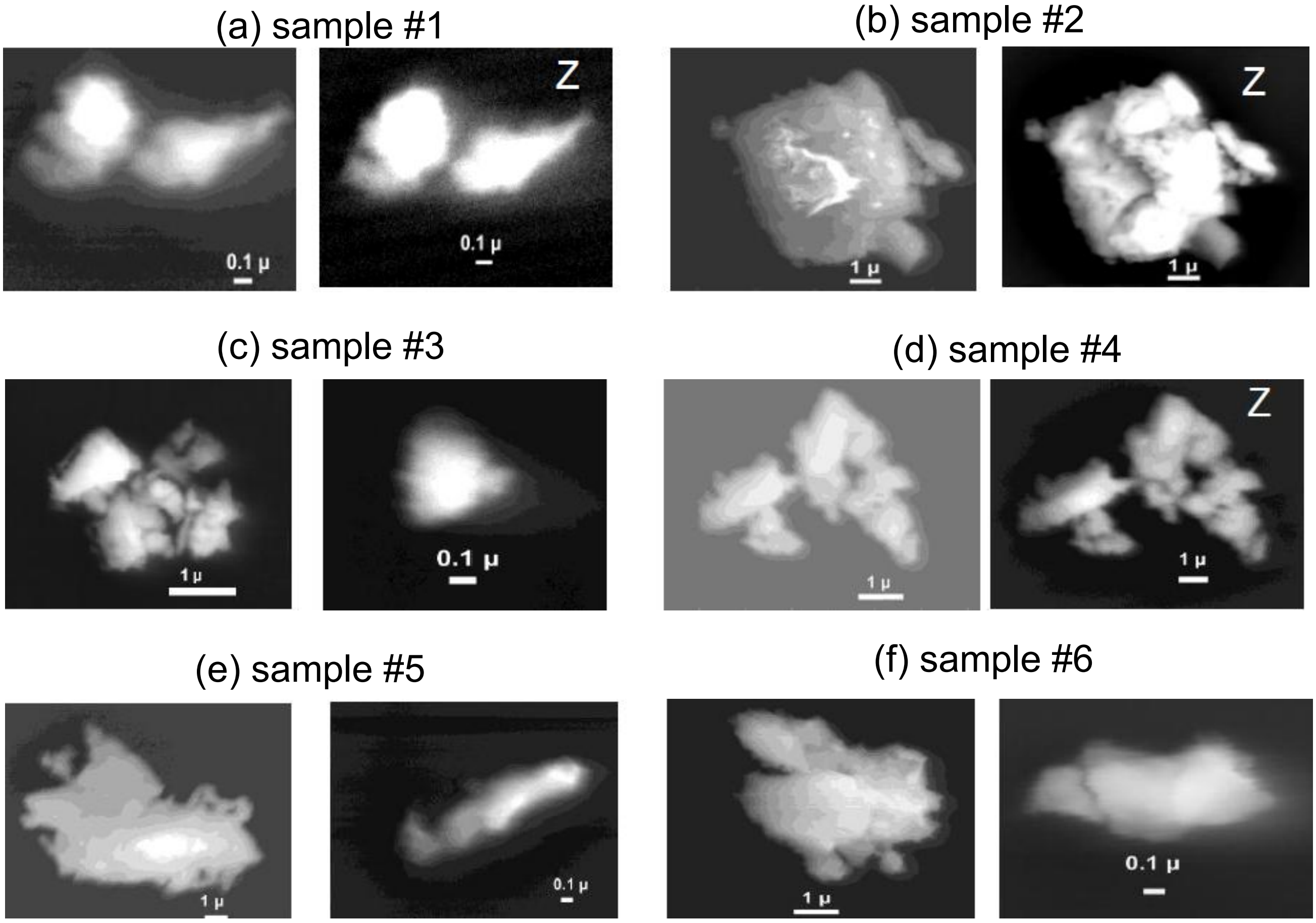

**Figure 2.** SEM images of $Hf_{0.5}Zr_{0.5}O_2$ nanopowders annealed at temperatures 500 – 800 °C (samples #1 – 4), $ZrO_2$ (sample # 5) and $HfO_2$ (sample # 6) nanopowders annealed at 800 °C. Abbreviation “Z” is for the Z contrast mode of the SEM; other images correspond to the second electron mode.

### B. EPR spectroscopy of the $Hf_{0.5}$ $Zr_{0.5}O_2$ nanopowders

The EPR measurements were performed at 300 K on a commercial Bruker Elexsys E580 spectrometer operating at 9.8 GHz frequency (X-band) using the 100 kHz field modulation. The effective g factors were calculated using the equation: $g_{ef} = h\nu/\mu_B H_r$, where $h$ is the Planck constant, $\nu$ is the operating frequency, $\mu_B$ is the Bohr magneton, and $H_r$ is a magnetic resonant field. The fitting parameters of the spectra were calculated using the "EasySpin" software [32].

The EPR spectra of $Hf_{0.5}Zr_{0.5}O_2$ nanopowders annealed at different temperatures, $HfO_2$ and $ZrO_2$ nanoparticles are presented in **Fig. 3(a).** Note that the spectral intensities are not normalized and are chosen for the convenience of presenting the spectrum in one figure. The main features presented at the spectra are labelled S1-S4. The feature S1 has $g_{ef}$ around 4.3, that usually corresponds to $Fe^{3+}$ ions located in a distorted oxygen environment [33]. Such EPR signal usually arises from uncontrolled iron impurities present in samples. Considering the uncontrolled nature of iron impurities in the structures under study, it can be assumed that the feature S1 has approximately the same intensity in all samples, which allows us to judge the real relative intensity of the EPR signals in a series of studied compounds.

The broad line S2 with a g-factor in the range of 2.0 – 2.4 is characteristic of paramagnetic ions coupled via the dipole-dipole and super-exchange interactions [34]. It can originate from the sample area where the concentration of the paramagnetic ions is high enough to form clusters with a high degree of interaction between them. In $Hf_{0.5}Zr_{0.5}O_2$, in addition to the uncontrolled $Fe^{3+}$ impurity, the formation of paramagnetic ions $Hf^{3+}$ and $Zr^{3+}$ is also possible. Considering that both Hf and Zr have the oxidation state of +4 in this compound, and these ions have a closed electron shell, the formation of $Hf^{3+}$ and $Zr^{3+}$ is possible due to the presence of an oxygen vacancy ($V_O$) or other defects near them. A large number of defects in the surface layer of nanoparticles contribute to the formation of regions with high concentration of paramagnetic ions and, most likely, the line S2 originates from the surface layer of the powders. The intensity of this line also decreases with an increase in annealing temperature suggesting that the number of paramagnetic defects decreases due to formation of less defective crystalline structure.

The sharp line S3 has g factor of 2.0031, and the peak-to-peak linewidth of 4.5 – 6.5 G. Its intensity decreases sharply with an annealing temperature increase. The residual carbon- and nitrogen-contained inclusions and paramagnetic radicals (present in the samples as a result of synthesis) could give a line with such g factor value [35]. This line does not saturate in intensity with increasing of the microwave power that suggests high concentration of such defects that leads to exchange narrowing of this line. Increase in the annealing temperature carried out in an air atmosphere leads to a decrease in volume of the carbon- and nitrogen-contained inclusions and vacancies that results in the decrease in the intensity of the line S3. This well agrees with X-ray diffraction data.

The feature S4 has a complex structure shown in **Fig. 3(b)** for the EPR spectra in a narrow range. The spectrum of the $HfO_2$ nanoparticles can be decomposed into three lines. Line 1 corresponds to the axial g-tensor with $g_{\parallel}$ = 1.938 and $g_{\perp}$ = 1.972, which is in good agreement with literature data obtained for $Hf^{3+}$ ions in a monoclinic $HfO_2$ [36]. The lines 2 and 3 with $g_x$ = 2.000, $g_y$ = 1.999, $g_z$ = 1.998 and $g_x$

= 2.046, $g_y$ = 2.064, $g_z$ = 2.079 respectively, which are present in the EPR spectra of $HfO_2$ samples studied in Ref. [36], can be assigned to oxygen vacancies with one trapped electron and to a hole-type oxygen centers. The spectrum of $ZrO_2$ nanoparticles in addition to lines corresponding to $Zr^{3+}$ ion with $g_{\parallel}$ = 1.957 and $g_{\perp}$ = 1.976 and assigned to oxygen-related defects [36, 37] also contains the line 4 with a small intensity at g = 2.003, corresponding to radicals.

The EPR spectra of the $Hf_{0.5}Zr_{0.5}O_2$ nanoparticles annealed at different temperatures can be well fitted by a superposition of the above-mentioned lines. In the spectrum of the sample #2 annealed at 600°C, the line S3 still dominates, but the spectrum already shows features at g = 1.946 – 1.975 corresponding to isolated not exchange coupled $(Hf/Zr)^{3+}$ centers (see **Fig. 3(b)**). The spectrum in the sample #3 annealed at 700°C can be well fitted with the orthorhombic symmetry g-tensor $g_x$ = 1.975, $g_y$ = 1.976, $g_z$ = 1.946, and the sample #4 annealed at 800° C can be fitted with $g_x$ = 1.975, $g_y$ = 1.972 and $g_z$ = 1.946. The almost coinciding g-tensor values $g_x$ and $g_y$ indicate that the local symmetry of the paramagnetic center is close to axial due to the presence of oxygen vacancy that produces strong perturbation along the Hf/Zr – $V_O$ direction. However, the absence of clearly defined features on the high-field side of the spectra, especially for the sample #3 annealed at 700°C, makes the determination of g values less reliable.

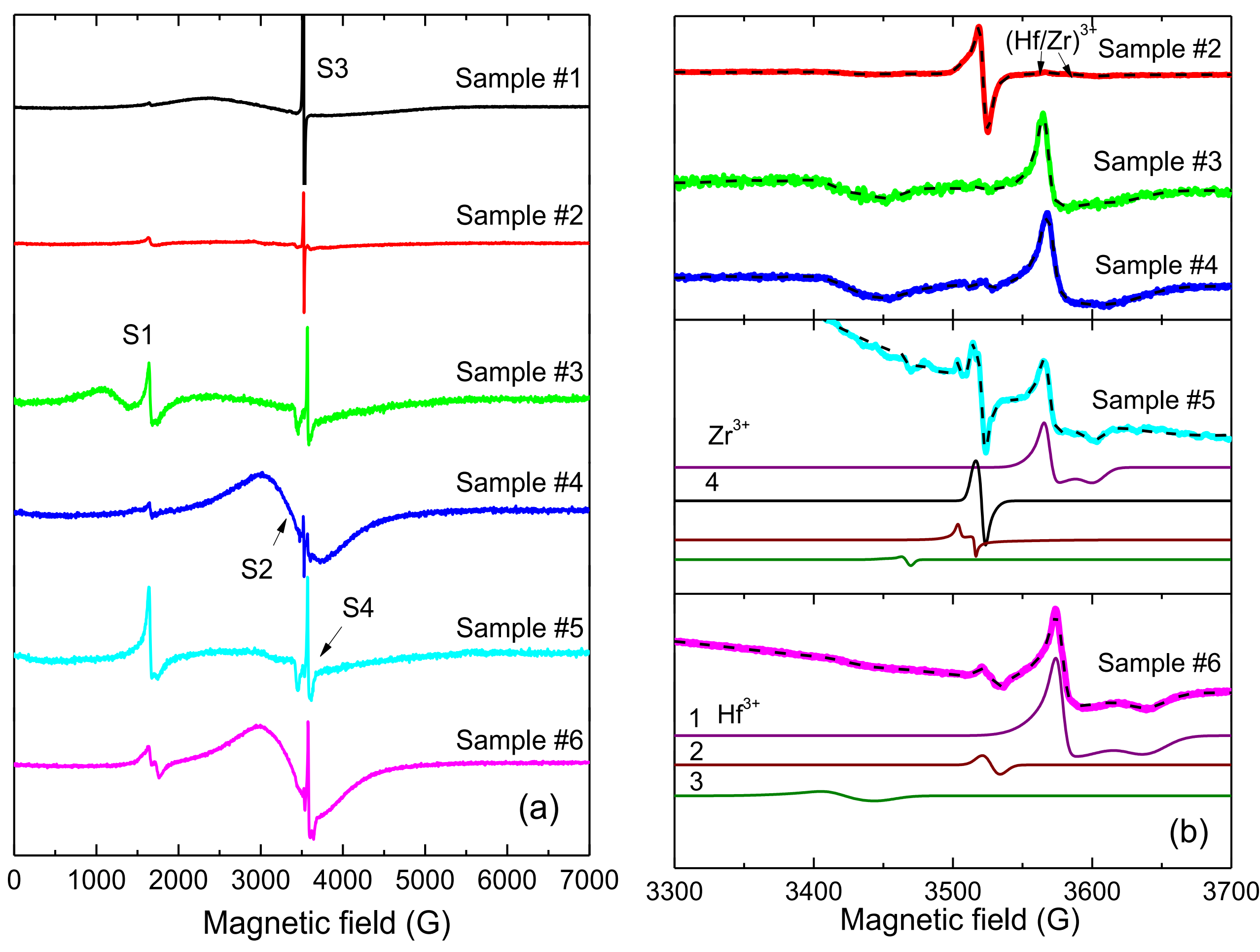

**Figure 3. (a)** The EPR spectra of $Hf_{0.5}Zr_{0.5}O_2$ nanopowders annealed at 500 – 800 °C (samples #1 – 4), $ZrO_2$ and $HfO_2$ nanopowders (samples #5 and #6, respectively). **(b)** Detailed EPR spectra of $Hf_{0.5}Zr_{0.5}O_2$, $HfO_2$ and $ZrO_2$ nanopowders (solid curves) and their fitting by the suggested centers $Zr(Hf)^{3+}$-$V_O$, $V_O$ – $e^-$ ($F^+$) (dashed black curves). The spectral lines1-4 are described in the text.

According to the literature, bulk hafnium and zirconium oxides exist in orthorhombic phases only at high pressures (see e.g., Refs. [38, 39]). There are several papers reporting about $Hf_{1-x}Zr_xO_2$ nanoparticles in the orthorhombic phase (see e.g., Refs. [19, 20, 21] and refs. therein), but we could not find any paper, which includes the EPR studies either for these particles or for bulk materials in the orthorhombic phase. However, according to the literature, there are the EPR measurements of $ZrO_2$ nanoparticles in the tetragonal phase [40], where the g-tensor of $Zr^{3+}$ has a definite axial symmetry, while our measurements show the g-tensor with rather orthorhombic symmetry.

The above analysis allows us to conclude that the local symmetry of the paramagnetic centers in $Hf_{0.5}Zr_{0.5}O_2$ annealed at higher temperatures is mostly in a monoclinic or a mixture of monoclinic and tetragonal phases, which is consistent with the X-ray diffraction data obtained for the samples #4-6 annealed at 800°C. Moreover, the decrease in the amount of the orthorhombic phase with an increase in the annealing temperature, according to XRD data, correlates with a decrease in the intensity of spectral lines associated with impurities and oxygen vacancies. This fact indicates the participation of oxygen vacancies and other defects in the formation of the orthorhombic phase in the $Hf_{0.5}Zr_{0.5}O_2$ powders.

### C. Electrophysical measurements

To study the transport of electric charge carriers in $Hf_xZr_{1-x}O_2$ nanopowders, the following sample preparation was used. The tableted powder sample was placed in a Teflon cylinder between two brass plungers, which serve to create uniaxial pressure and also are used as electrical contacts (see **Appendix C** for details). The distance between the plungers was determined by the thickness of the sample. The diameter of the samples was 4 mm, the distance between the contacts was (100 ± 10) μm. The voltage sweep across the sample was carried out by the software-controlled Instek PSP 603 power supply with a step of 20 mV every 2 s. The recording of the voltage drop across the sample and the load resistor was performed every 2 seconds by Keithley 2000-SCAN precision multimeter. The sweep time was selected in such a way to avoid the influence of transient processes on the measurement of the I-V characteristic. The total measurement time for one sample in the DC mode was more than 4000 s. Measurements of capacitance of the studied samples in the range ($4 - 5\cdot10^5$) Hz were carried out by the RLC-meter LCX200 ROHDE & SCHWARZ at the temperature of 290 K. The samples annealing temperature, phase composition and main features of their electrophysical properties are given in **Table 1** and illustrated in **Figure 5**.

**Table 1.** Phase composition and electrophysical properties of the $Hf_xZr_{1-x}O_2$ nanopowder samples

| **Sample number** | # 1 | # 2 | # 3 | # 4 | # 5 | # 6 |
|---|---|---|---|---|---|---|
| **Composition** | $Hf_{0.5}Zr_{0.5}O_2$ | $Hf_{0.5}Zr_{0.5}O_2$ | $Hf_{0.5}Zr_{0.5}O_2$ | $Hf_{0.5}Zr_{0.5}O_2$ | $ZrO_2$ | $HfO_2$ |
| **Annealing temperature** (°C) | 500 | 600 | 700 | 800 | 800 | 800 |
| **Monoclinic phase (vol. %)** | 15 ± 4 | 9 ± 1 | 58 ± 3 | 93 ± 2 | 84 ± 5** | 100 ± 2 |
| ***Orthorhombic phases (vol. %)** | 85 ± 11 | 91 ± 4 | 42 ± 5 | 7 ± 1 | 0 | 0 ± 2 |
| ****Tetragonal phase (vol. %)** | 0 | 0 | 0 | 0 | 16± 5 | 0 |
| **NDR region (voltage region)** | highly resistive | exist (0 – 2.5 V) | absent | absent | absent | exist (0 – 2.0 V) |
| **Resistive I-V hysteresis loop** | highly resistive | very slim loop | slim loop | pronounced loop | pronounced loop | slim loop |
| **Capacitance at 4 Hz (pF)** | 18 at 0.1 V<br>13 at 5.0 V | 360 at 0.1 V<br>150 at 5.0 V | 545 at 0.1 V<br>200 at 5.0 V | 108 at 0.1 V<br>44 at 5.0 V | 800 at 0.1 V<br>380 at 5.0 V | 2500 at 0.1 V<br>1050 at 5.0 V |
| **Dielectric permittiviy at 4 Hz** | 19.4 | 388 | 587 | 116 | 947 | 2691 |

*As determined from XRD analysis, we observe three orthorhombic phases and one monoclinic phase.

** As determined from XRD analysis, only the $ZrO_2$ sample # 5 contains about 16 vol.% of tetragonal phase.

**Figure 4(a)** shows the fraction of the orthorhombic phases at first to increase slightly and then decrease strongly with an increase of the annealing temperature from 500°C to 800°C. The fraction of the monoclinic phase, at first decreases slightly and then increases strongly with an increase of the annealing temperature from 500°C to 800°C. **Figure 4(b)** shows the dependence of the dielectric permittivity (measured at 4 Hz) and resistivity (measured in the AC and DC modes) versus the annealing temperature. The AC resistivity was measured at the voltage amplitude of 500 mV and frequency of 4 Hz. The DC resistivity was determined from the I-V curves measured at the applied voltage of 10 V. It is seen that all parameters except of the DC resistivity manifest a nonmonotonic dependence on the annealing temperature.

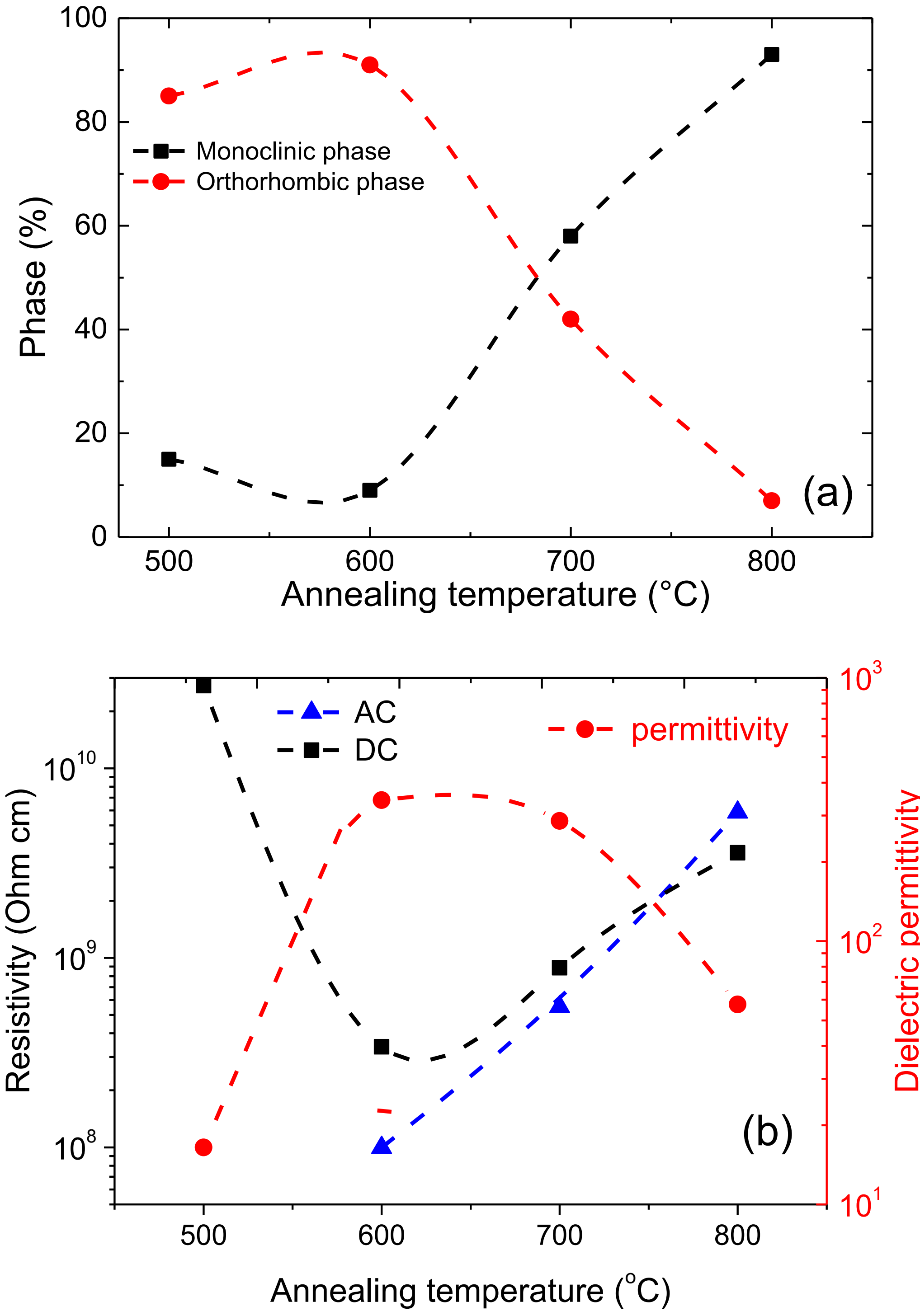

**Figure 4. (a)** The fraction of monoclinic and orthorhombic phases versus the annealing temperature of the $Zr_{0.5}Hf_{0.5}O_2$ nanopowders (samples #1 – 4). **(b)** The dielectric permittivity (red circles, the right scale) and resistivity (black squares and blue triangles, the left scale) versus the annealing temperature. Measurements of AC resistivity and dielectric permittivity are performed at 4 Hz.

**Figure 5** shows the dependence of the dielectric permittivity (measured at 4 Hz) and resistivity (measured in AC, $f$ =4 Hz, and DC modes) versus the fractions of the orthorhombic and the monoclinic phases. It is seen that all parameters except of the DC resistivity manifest a complicated dependence on the phase fraction. The DC-resistivity monotonically increases with increasing the monoclinic phase fraction, while the AC-resistivity has a pronounced maximum in the sample #1 (annealed at 500°C) where, as shown above, crystallization only begins, and thus the temperature of 500°C is insufficient for burning out the remainders of the organic phase.

It is worth noting that a sharp maximum in the AC resistivity and the minimum in the dielectric permittivity curve are not expected from the supposed correlated change in the ratio of the orthorhombic and the monoclinic phases, which occur with increase in the annealing temperature. However, in accordance with the XRD data, one can suppose that the deviation at the annealing temperature of 500ºC is not related to the ratio of phases, but rather is caused by noticeable fraction of the amorphous phase forming at the very beginning of the crystallization process. Therefore, it looks reasonable to exclude the point at 500ºC and then to relate the dependence of both, the AC resistivity and the dielectric permittivity, vs the ratio phases. These smooth and well-correlated behavior is shown in **Fig. 5** by the dash-dotted extrapolation curves.

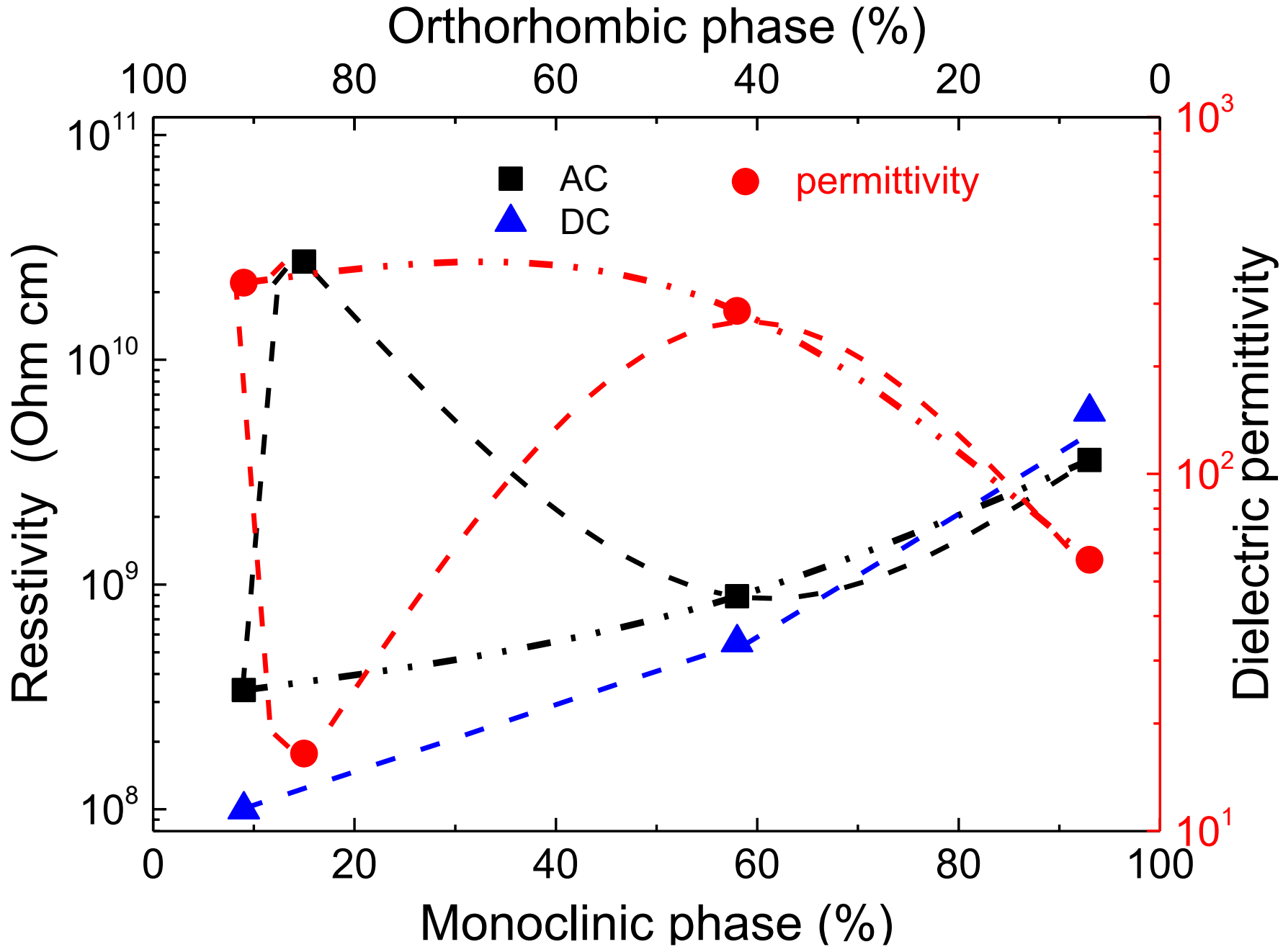

**Figure 5.** The dielectric permittivity (red circles, the right scale) and resistivity (black squares and blue triangles, the left scale) of the $Zr_{0.5}Hf_{0.5}O_2$ nanopowders (samples #1 – 4) versus the fractions of orthorhombic (the upper scale) and monoclinic (the lower scale) phases. Measurements of AC resistivity and dielectric permittivity are performed at 4 Hz. The dash-dotted curves are the interpolation dependences of the AC resistivity and dielectric permittivity vs the ratio of orthorhombic and monoclinic

phases without considering the data point for the sample annealed at 500°C. The dashed curves are the interpolations, which include the data for the sample annealed at 500°C.

The sample #1 has a very high resistivity, which value is beyond the measuring range. The results of measurements of the I-V curves of the $Hf_{0.5}Zr_{0.5}O_2$, $ZrO_2$ and $HfO_2$ nanopowders (samples # 2-6) are presented in **Fig. 6**. A pronounced asymmetry in the forward and backward branches of the I-V curve for the sample #2, as well as less pronounced asymmetry of the I-V curves for samples #3-6, can be associated with the effect of the charge accumulation by the surface of the nanoparticles. For comparative analysis, we consider further the results of the I-V curves measurement corresponding only to the forward bias polarization shown in **Fig. 7**, where the forward and backward voltage sweeps are shown in different colors.

The current magnitude through the $Hf_{0.5}Zr_{0.5}O_2$ nanopowder samples decreases at equal bias with an increase in the annealing temperature, while the difference in the currents between the forward and backward voltage sweep branches increases. This loop-like behavior is the manifestation of changes in their resistance, i.e., the resistive switching effect. The effect may occur due to either accumulation of charges on the opposite sides of a sample and nanoparticles shells in the case of switching to a higher resistivity, or formation of filaments with higher conductivity due to phase transformation when resistivity switches to a smaller magnitude [23]. We can relate the origin of the observed resistive switching effect with migration of oxygen vacancies (confirmed by EPR spectra analysis) and with the significant amount of the orthorhombic phase in the samples #2 and 3 (confirmed by XRD analysis). The migration of oxygen vacancies, which are elastic dipoles, can support the formation of polar nanoregions, which "trap" the space charge carriers due to the self-screening effect [41].

The change in the current behavior for the nanopowders annealed at different temperatures can be determined by different resulting concentrations of charge carriers independently of the ratio between Zr and Hf. So, for the samples # 5 and # 6, which are pure Zr and Hf oxides, the current values are comparable.

Note that the I–V curves obtained for samples # 2 and #6 reveal the region with the negative differential resistance (NDR), which is observed in the voltage range 1-2 V, corresponding to the electric field strength of 100 - 200 V/cm. The nature of the NDR effect becomes clear in the dependence of resistance on the applied voltage depicted in **Fig. 8**. It is seen that at first the resistance increases with increasing voltage, reaches a maximum and then slowly decreases with further increase in voltage. On the backward voltage sweep the resistance repeats the trace till the maximum of the forward sweep and then sharply increases keeping high value down to the lowest voltage magnitude. So, one can relate a possible explanation of the NDR appearance with the capture of carriers by traps via coming over or tunneling through a potential barrier at growing electric field.

In the backward voltage sweep, the electric current stays around zero at small voltages of about (1.4 − 1.7) V for all samples, which also evidences the charge accumulation in the samples. When the applied voltage crosses zero, a small negative current, i.e., with the opposite direction, is recorded in the circuit, which magnitude decreases with time, that also indicating that the charge accumulation occurs entire the sample volume.

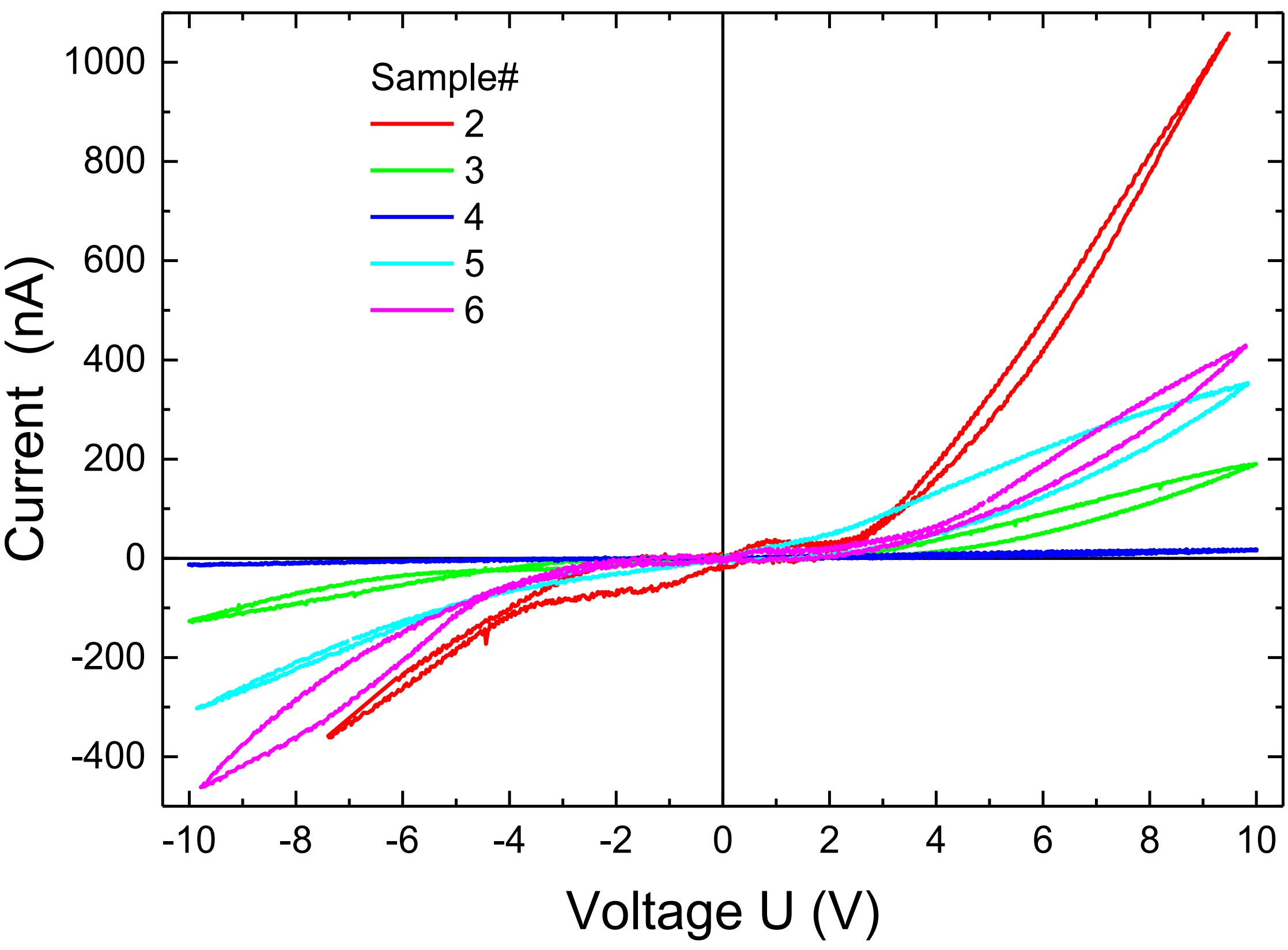

**Figure 6.** I-V curves of the $Hf_{0.5}Zr_{0.5}O_2$ nanopowders annealed at 600 – 800 °C (samples # 2 – 4), $ZrO_2$ and $HfO_2$ nanopowders annealed at 800°C (samples # 5 and # 6, respectively).

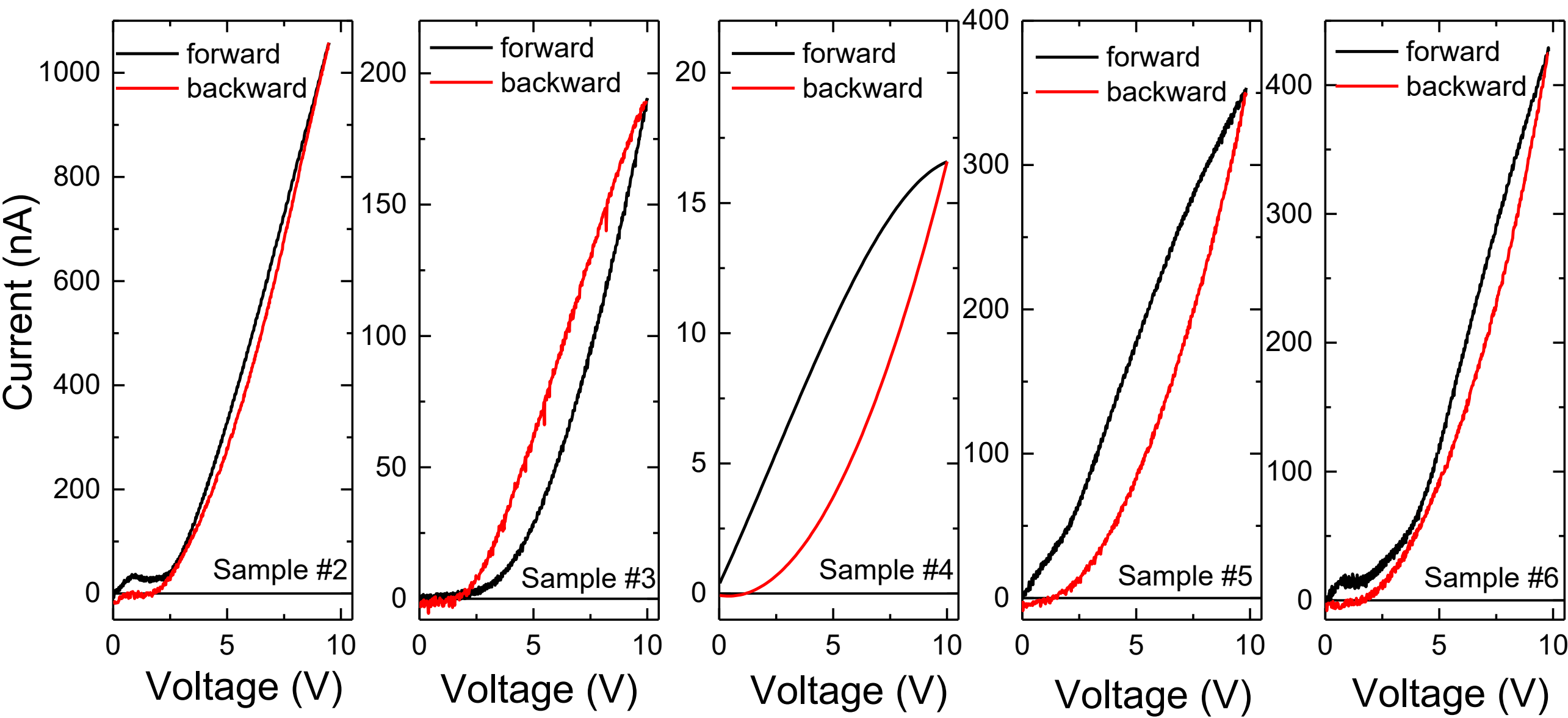

**Figure 7.** The I-V curves at forward bias for the $Hf_{0.5}Zr_{0.5}O_2$ nanopowders annealed at 600 – 800 °C (samples #2 – 4), $ZrO_2$ and $HfO_2$ nanopowders annealed at 800°C (samples # 5 and #6, respectively). Black and red curves correspond to the forward and backward direction of the voltage sweep.

To illustrate the resistive switching effect in the $Hf_{0.5}Zr_{0.5}O_2$ nanopowder samples, the dependences of resistance on the applied voltage are plotted in **Fig. 8** for the samples #2 - 4. It is seen that the branches corresponding to the forward and backward voltage sweeps almost coincide at the voltage values above 2.5 V in the sample #2 with the NDR effect. At smaller voltages the resistance corresponding to the forward and backward sweeps differs more than by order of magnitude. For the other two samples, the magnitude of the difference in the resistance values between the forward and backward sweeps increases with decrease in applied voltage.

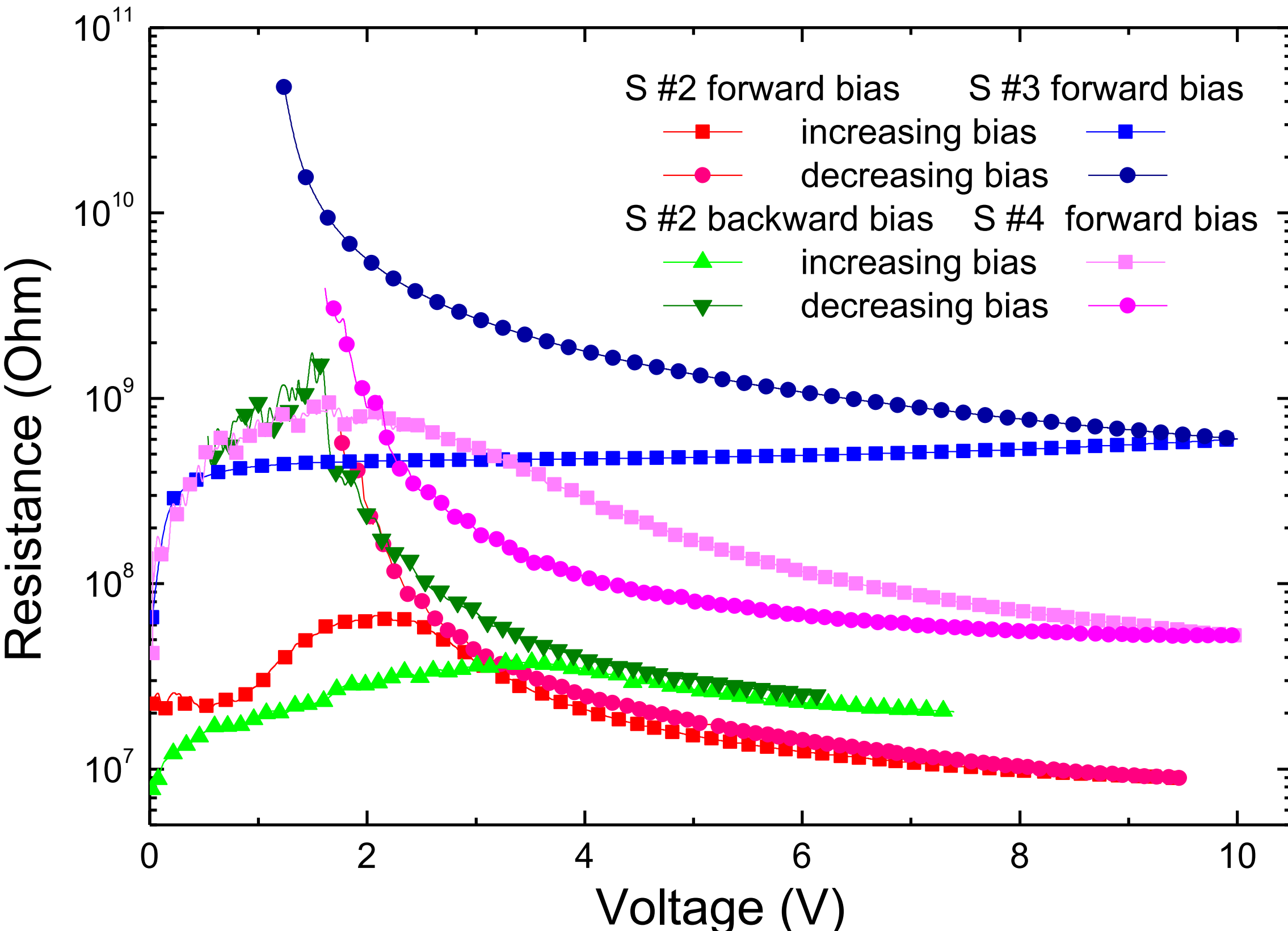

**Figure 8.** Resistance of the $Hf_{0.5}Zr_{0.5}O_2$ nanopowders (samples #2, #3 and #4) depending on the applied voltage.

**Figure C1** (in **Appendix C**) shows the frequency dependences of the capacitance and the loss angle tangent of the samples #1-6 measured at different amplitude of the sinusoidal voltage applied to the samples. The sample # 1 demonstrates a weak dependence of the capacitance vs. the frequency and the level of the applied ac voltage (from 0.1 to 5 V), which is the additional evidence that the sample structure is amorphous. A stronger dependence of the capacitance on the frequency and ac voltage is characteristic for the samples # 2-6. Also, a non-monotonic dependence of the dielectric loss angle tangent on the frequency is observed for these samples.

The frequency dependences of the effective dielectric permittivity and loss angle tangent for the nanopowder samples of $Hf_{0.5}Zr_{0.5}O_2$, $ZrO_2$ and $HfO_2$ are shown in **Fig. 9(a)** and **9(b)**, respectively. The permittivity of all samples decreases monotonically with an increase in frequency. The strongest difference between the curves is observed in the lower frequency range. At the same time, the features of the effective dielectric permittivity poorly correlate with the ratio of monoclinic and orthorhombic phases. Concerning the loss angle tangent, they may be divided into 2 groups. The first group of samples manifests a monotonous decrease with increasing frequency. The samples of this group have larger resistivity (see **Fig. 9(c)**). The second group of samples reveal a pronounced maximum in the middle of the frequency range and their resistivity is lower. At that, the dependences of resistivity versus frequency decrease monotonously for all samples.

The dielectric permittivity of all samples, except the sample #1, varies with increasing frequency by several (2-3) orders of magnitude. Some of the samples have a maximum in the frequency dependences of the loss tangent tgδ, or tendency to that. The resistivity versus frequency dependences manifest change of their slope at a certain frequency. These peculiarities may indicate the change in the conduction mechanisms with growing frequency [42], as well as be related to the dipolar (i.e., polarization) contribution. Notably that the sample #2 has the highest content of the orthorhombic phase in accordance with the XRD analysis.

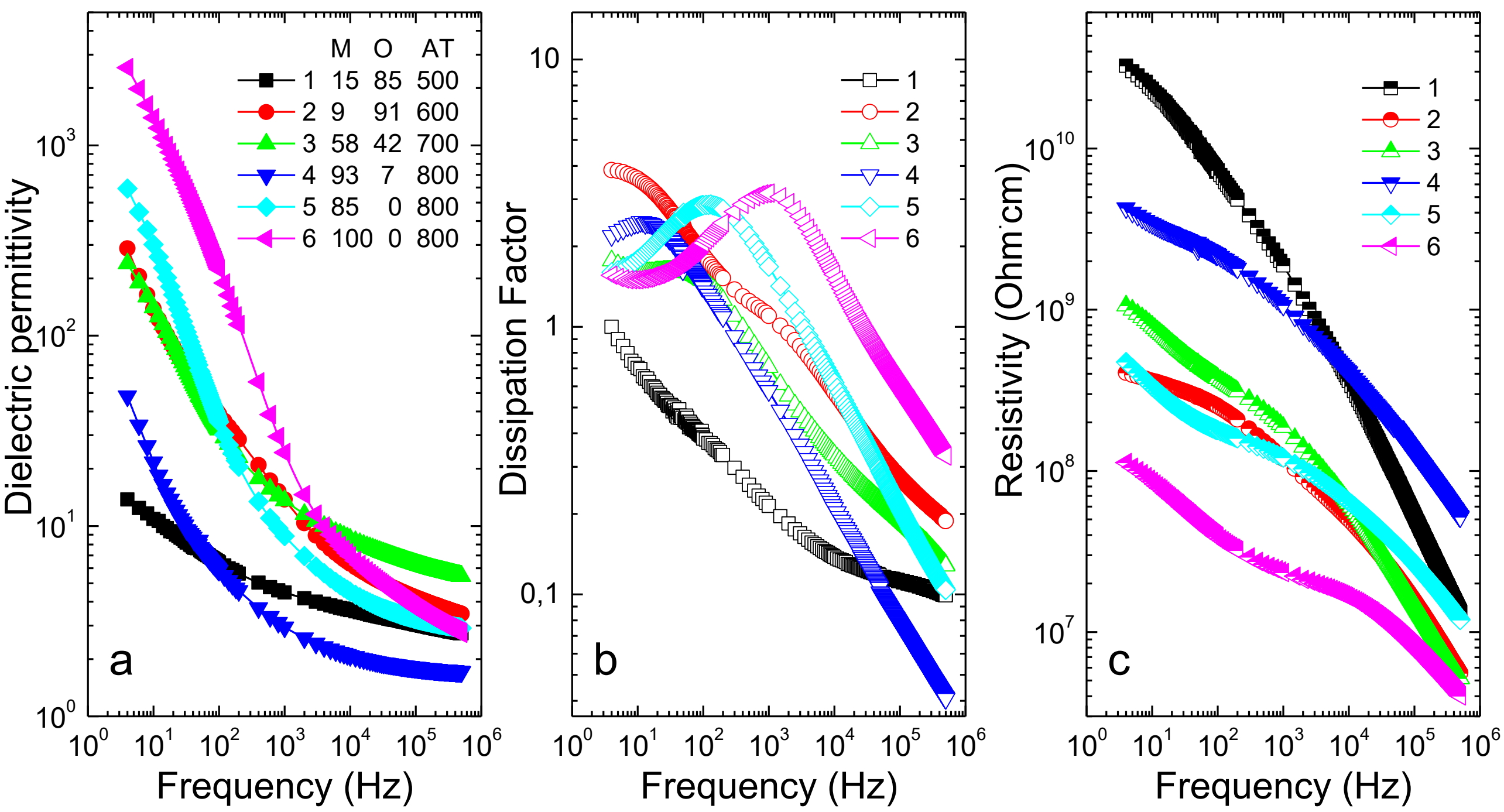

**Figure 9.** Frequency dependences of effective dielectric permittivity ε **(a),** the loss tangent angle **(b)** and ac resistivity for the $Hf_{0.5}Zr_{0.5}O_2$ nanopowders (samples #1 – 4), $ZrO_2$ nanopowder (sample # 5) and $HfO_2$ nanopowder (sample #6). The ac voltage amplitude is 0.5 V. The samples are marked by the percentage content of the monoclinic (M) and orthorhombic (O) phases and annealing temperature (AT).

One should note the correlation between the values of the effective dielectric permittivity in the $Hf_{0.5}Zr_{0.5}O_2$ powders and their annealing temperature which is illustrated in **Fig. 10**. It is seen that the permittivity has a diffuse maximum in the annealing temperature range (600 – 700) °C for all frequencies. At the same time, it poorly correlates with the ratio of monoclinic and orthorhombic phases. However, the $Hf_{0.5}Zr_{0.5}O_2$ powders, which reveal the NDR region in the I-V curves, possess both a high content of the orthorhombic phase and a high permittivityThis fact supports the hypothesis that ferroelectric-like polar regions could exist in the samples.

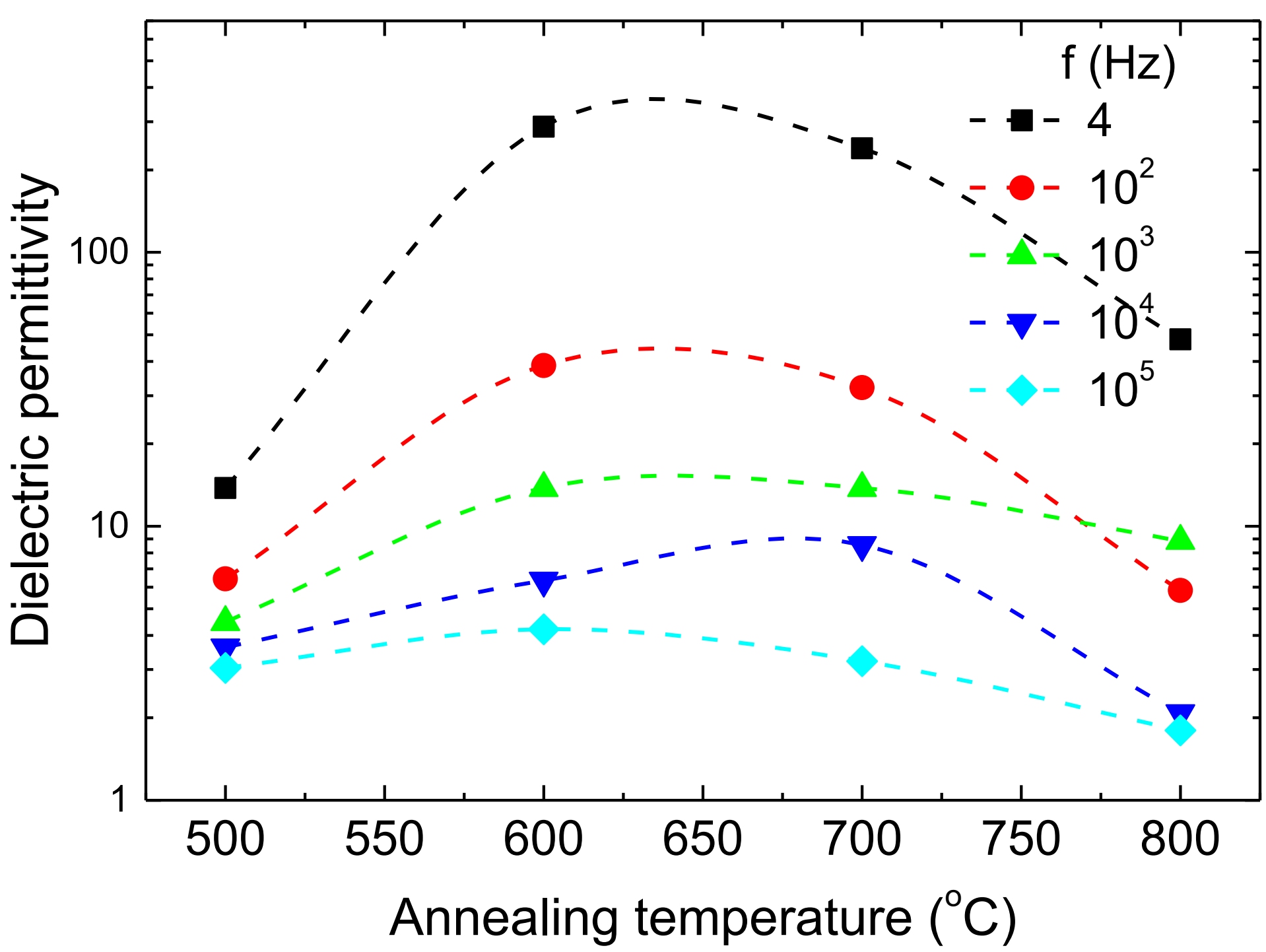

**Figure 10.** Dependence of the permittivity of the $Hf_{0.5}Zr_{0.5}O_2$ samples on the annealing temperature at different frequencies.

To summarize the section, let us underline that all frequency dependences show a certain correlation with I-V curves with respect to the annealing temperature. The whole set of results for $Hf_{0.5}Zr_{0.5}O_2$ samples indicates on possibility to realize the NDR region in the annealing temperatures range 600 – 700°C through obtaining a higher content of the orthorhombic phase in the hafnia-zirconia nanoparticles.

## THEORETICAL DESCRIPTION

### A. "Effective" Landau-Ginzburg-Devonshire model

Analysis of the spatial-temporal evolution of polarization in the $Hf_xZr_{1-x}O_2$ nanoparticles covered by ionic-electronic charge with a surface charge density $\sigma_S$, which appear due to the surface adsorption of electrochemically active species (ions and/or vacancies), is performed in the framework of "effective" LGD model [21 - 22]. This approach incorporates elements of the Kittel-type model [43], incorporating both polar and antipolar modes [44, 45, 46], with the Landau theory with effective parameters [47] extracted from the experiment [9]. Within this model, the free energy functional $F$ is expressed as the sum of several key terms:

$$F = F_{bulk} + F_{grad} + F_S. \quad (1)$$

The first term $F_{bulk}$ is the bulk free energy, that is a series expansion in the second and fourth powers of the polar ($P_f$) and antipolar ($A_f$) order parameters:

$$F_{bulk} = \int_{V_f} d^3r \left(\frac{a_P}{2}P_f^2 + \frac{b_P}{4}P_f^4 + \frac{\eta}{2}P_f^2 A_f^2 + \frac{a_A}{2}A_f^2 + \frac{b_A}{4}A_f^4 - E_f P_f\right). \quad (2)$$

Here $V_f$ is the volume of the $Hf_xZr_{1-x}O_2$ nanoparticle. The coefficients $a_P$ and $a_A$ can be strain-dependent as shown in Refs. [48, 49]. The polar and antipolar parameters gradient energy $F_{grad}$ and the surface energy $F_S$ are listed in Refs.[21 - 22]. Electric field $E_f$ obeys the Poisson equation considering the presence of surface ionic-electronic charges. Calculation details are given in Refs. [12, 14, 21-22].

The Stephenson-Highland model [50] describes the relationship between the surface charge density $\sigma_S[\delta\phi]$ and electric potential excess $\delta\phi$ at the nanoparticle surface. This model accounts for the coverages of positive ($i = 1$) and negative ($i = 2$) surface charges (e.g., ions) in a self-consistent manner. The corresponding Langmuir adsorption isotherm [51, 52] for the charge density $\sigma_S$ is given by expression:

$$\sigma_S[\delta\phi] \cong \sum_i \frac{eZ_i}{A_i}\left(1 + a_i^{-1} \exp\left[\frac{\Delta G_i + eZ_i\delta\phi}{k_B T}\right]\right)^{-1}, \quad (3)$$

where $e$ is the electron charge, $Z_i$ is the ionization number of the adsorbed ions, $a_i$ is the dimensionless chemical activity of the ions in the environment (as a rule $0 \leq a_i \leq 1$)), $T$ is the absolute temperature, $A_i$ is the area per surface site for the adsorbed ion, and $\Delta G_i$ are the formation energies of the surface charges (e.g., ions and/or electrons) at normal conditions; the subscript $i$ =1, 2. Mismatch and/or chemical strains exist at the nanoflake-substrate interface. From Eq.(5), the effective screening length $\lambda$ associated with the adsorbed charges is $\frac{1}{\lambda} \approx \sum_i \frac{(eZ_i)^2 a_i \exp\left[\frac{\Delta G_i}{k_B T}\right]}{\varepsilon_0 k_B T A_i\left(a_i + \exp\left[\frac{\Delta G_i}{k_B T}\right]\right)^2}$ [52].

To calculate an electric current in the system with resistive switching, a classical theory of memristors [53] and their interconnected networks [54] can be used. The basic formulation of memristive switching theory can be found in Strukov et al. papers [55, 56, 57], who revealed that the memristive-type behavior is inherent to thin films with mixed ionic-electronic conductivity, when the drift-diffusion kinetic equations for electrons, holes and mobile donors/acceptors are non-linearly coupled, and thus the resistance "memory" depends on the thickness ratio of the doped and pure regions of the film. Later, it was shown that the space charge dynamics considering steric effect and chemical stresses can lead to the resistive switching and ferroelectric-like hysteresis of electromechanical response in oxide films with impurity ions and/or oxygen vacancies [58]. Below we use the theoretical model [58] to explain the behavior of the charge-voltage (Q-V) and current-voltage (I-V) curves in $Hf_{0.5}Zr_{0.5}O_2$ nanoparticles with adsorbed ionic-electronic charge. We also assume that the ionic-electronic charge density $\sigma_S$ depends strongly on the annealing conditions, being closely related to the concentration of oxygen vacancies, which accumulation appears at the particle surface and under the surface (due to the strong reduction of the vacancy formation energy in the vicinity of surface).

### B. Modelling results and discussion

We revealed that the ferro-ionic coupling, originated from the interaction of the surface charges (e.g., impurity ions and/or oxygen vacancies), can induce the ferroelectric-like polar regions charge accumulation in the $Hf_{0.5}Zr_{0.5}O_{2-y}$ nanoparticles at relatively small density of the surface charge $\sigma_S$. The ferro-ionic coupling can induce and/or support the long-range polar order at higher $\sigma_S$ [21-22].

As it was shown by DFT (see calculation details in Ref. [19]), the difference between the energies of the ferroelectric-like orthorhombic phases and the monoclinic phase calculated in the $HfO_2$ structure by the LDA approximation is rather small (~20 meV/f.u.). The density functional theory (DFT) calculations were performed using the full-potential all-electron local-orbital (FPLO) code [59] for the local density approximation (LDA) functional (see **Appendix D** for details). The calculated equilibrium lattice parameters $a$, $b$, and $c$ were compared with the experimental results [60, 61, 62] and with the values determined by reference DFT packages (see **Table D1** in **Appendix D**). As expected, the crystal structure with the monoclinic symmetry has the lowest ground state energy in the bulk $HfO_2$, $Hf_{0.5}Zr_{0.5}O_2$ and $ZrO_2$. At the same time, the difference between the energies of the ferroelectric orthorhombic phases (f-phase) and the monoclinic phase (m-phase) calculated in the $Hf_{0.5}Zr_{0.5}O_2$ structure appeared smaller than that determined by other DFT methods. The energy difference $\Delta U$ between the m-$Hf_{0.5}Zr_{0.5}O_2$ and f-$Hf_{0.5}Zr_{0.5}O_2$ ($\Delta U \approx 7.8$ meV/f.u.) is significantly smaller than the difference between the m-$HfO_2$ and f-$HfO_2$ ($\Delta U \approx 11.6$ meV/f.u.). Thus, the addition of Zr may stabilize the f-phase. Thus, the DFT calculations indicate that small $Hf_{0.5}Zr_{0.5}O_{2-y}$ nanoparticles may become polar, especially in the presence of impurity atoms and/or oxygen vacancies.

Polar clusters inside the spherical $Hf_{0.5}Zr_{0.5}O_2$ nanoparticles annealed at different temperatures $T_a$ are shown schematically by the grey color contrast in **Fig. 11(a)-(d)**. The abbreviation O/M corresponds to the ratio of orthorhombic to monoclinic phase fractions, which was used in the calculations of polar regions distributions (see details in Ref. [22]). Despite the O/M ratio is high for $T_a = 500$°C, any correlation of polar-like orthorhombic phase regions is absent in the image **(a)** due to the presence of conductive carbon inclusions, which act as conductive regions in the calculations. Polar regions appear in the form of labyrinthine domains for high O/M ratios (which corresponds to $T_a = 600$°C), shirk to the central part of the nanoparticles with decrease in the O/M ratio (which corresponds to $T_a = 700$°C), and almost disappear for small O/M ratio (which corresponds to $T_a = 800$°C) (compare the images **(b)**, **(c)** and **(d)** in **Fig. 11**).

The Q-V and I-V dependences, calculated for the 30-nm $Hf_{0.5}Zr_{0.5}O_2$ nanoparticles, annealed at different temperatures, are schematically shown in **Figs. 11(e)-(h)**. The transitions from the paraelectric-like to the ferroelectric-like Q-V loops, then to the antiferroelectric-like and dielectric-like Q-V loops occur with the increase in the $T_a$ from 500ºC to 800ºC. Corresponding I-V curves, which contain hysteresis regions, reveal the features of resistive switching and charge accumulation, which we relate with the possible appearance of the ferroelectric-like polar regions.

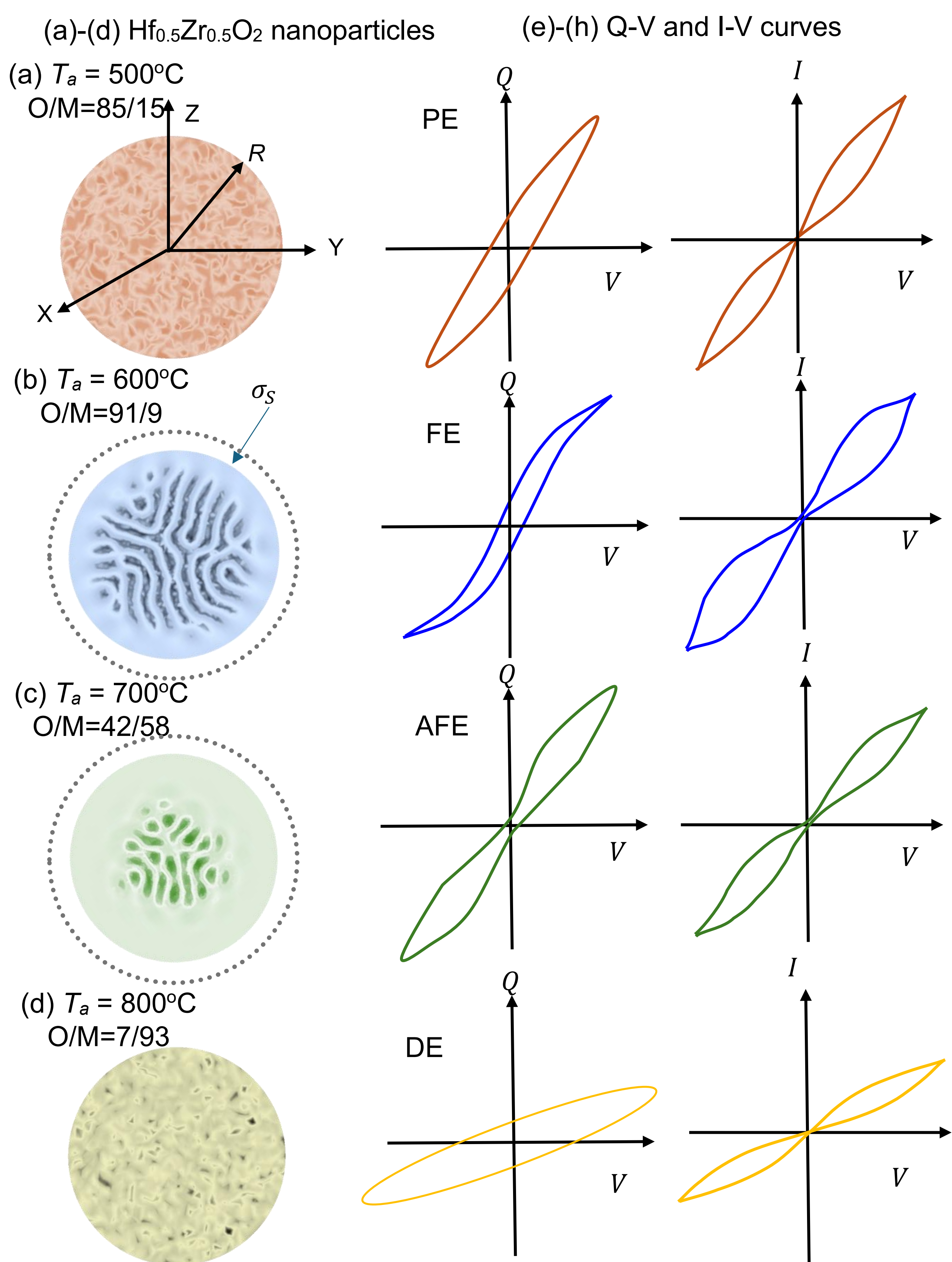

**Figure 11. (a)-(d)** Schematic picture of polar clusters inside the spherical $Hf_{0.5}Zr_{0.5}O_2$ nanoparticles annealed at different temperatures $T_a$. The shell of ionic-electronic screening charges (e.g., impurity ions or vacancies) with the density $\sigma_S$ is shown dotted circle. **(e)-(h)** The charge-voltage (Q-V) and current-

voltage (I-V) curves, calculated for the $Hf_{0.5}Zr_{0.5}O_2$ nanoparticles. The paraelectric-like (PE), ferroelectric-like (FE), antiferroelectric-like (AFE), and dielectric-like (DE) polarization loops are shown.

## CONCLUSIONS

We revealed the resistive switching and charge accumulation effect in $Hf_{0.5}Zr_{0.5}O_2$ nanopowders sintered by the auto-combustion sol-gel method and annealed at temperatures from 500°C to 800°C. The fraction of the orthorhombic phase, determined by the XRD, decreases from 91 vol.% to 7 vol.% with an increase in annealing temperature from 600°C to 800°C.

The EPR spectra reveal the great amount of oxygen vacancies in the annealed samples, at that the decrease of the orthorhombic phase fraction, which appears with an increase in the annealing temperature, correlates with a decrease in the intensity of EPR spectral lines associated with impurities and oxygen vacancies. This indicates the participation of oxygen vacancies and other defects in the formation of the orthorhombic phase in the $Hf_{0.5}Zr_{0.5}O_2$ powders.

The frequency dependences of resistance, capacitance and losses show a certain correlation with the current-voltage double loops with respect to the annealing temperature. The whole set of results for $Hf_{0.5}Zr_{0.5}O_2$ nanopowders indicates on possibility to realize the NDR region in I-V characteristics for the samples annealed at temperatures (600 – 700)°C. The temperature range corresponds to the highest content of the orthorhombic phase in the hafnia-zirconia nanoparticles.

The analysis allows us to relate the resistive switching, NDR and charge accumulation observed in $Hf_{0.5}Zr_{0.5}O_2$ nanopowders with the possible appearance of ferroelectric-like polar regions in the orthorhombic phase in the $Hf_{0.5}Zr_{0.5}O_2$ nanoparticles, which agrees with the calculations performed in the framework of LGD approach and DFT. In particular, the performed DFT calculations reveal that small $Hf_xZr_{1-x}O_2$ nanoparticles may become polar, especially in the presence of impurity atoms and/or oxygen vacancies.

## Authors' contribution

The research idea belongs to A.N.M., I.V.F. and O.S.P. O.S.P., V.V.V. and J.M.G. performed electrophysical measurements and interpreted results. I.V.F. and V.N.P. sintered the samples and characterized them by TDA and XRD measurements jointly with A.O.D. and M.M.K. V.I.S. performed SEM characterization. I.V.K. performed DFT calculations. E.A.E. wrote the codes and performed numerical calculations. Y.O.Z., L.P.Y. and V.V.L. performed EPR measurements. A.N.M. performed analytical calculations and wrote the manuscript draft jointly with O.S.P. and V.V.V. All co-authors discussed and analyzed the results, and corresponding authors made all improvements in the manuscript.

### Acknowledgments

Authors gratefully acknowledge useful discussion with Dr. Alexander S. Baskevich (Oles Honchar Dnipro National University). The work of O.S.P., J.M.G. and A.N.M. are funded by the National Research Foundation of Ukraine (grant N 2023.03/0132 "Manyfold-degenerated metastable states of spontaneous polarization in nanoferroics: theory, experiment and perspectives for digital nanoelectronics"). The work of E.A.E., I.V.K., Y.O.Z., A.O.D., M.M.K, and M.D.V. are funded by the National Research Foundation of Ukraine (grant N 2023.03/0127 "Silicon-compatible ferroelectric nanocomposites for electronics and sensors"). A.N.M. also acknowledges the support (materials characterization) from the Horizon Europe Framework Programme (HORIZON-TMA-MSCA-SE), project № 101131229, Piezoelectricity in 2D-materials: materials, modeling, and applications (PIEZO 2D). V.V.V. acknowledges the Target Program of the National Academy of Sciences of Ukraine, Project No. 5.8/25-П "Energy-saving and environmentally friendly nanoscale ferroics for the development of sensorics, nanoelectronics and spintronics"

### APPENDIX A. Synthesis of $Hf_{0.5}Zr_{0.5}O_2$ nanoparticles

The synthesis was carried out by the sol-gel method in the presence of citric acid as a gelling agent. The hafnium and zirconium sources were the crystal hydrates hafnyl nitrate $HfO(NO_3)_2{\cdot}xH_2O$ and zirconyl nitrate $ZrO(NO_3)_2{\cdot}yH_2O$, respectively ("x" and "y" correspond to the amount of $H_2O$ molecules). Preliminary thermogravimetric studies, which were carried out at 800°C for 50 minutes, allowed us to determine the exact composition of the starting components: $HfO(NO_3)_2{\cdot}2{,}5H_2O$ and $ZrO(NO_3)_2{\cdot}4{,}6H_2O$.

To obtain approximately 1 g of $Hf_{0.5}Zr_{0.5}O_2$ nanopowder, 0.9453 g of $ZrO(NO_3)_2{\cdot}4{,}6H_2O$ and 1.0905 g of $HfO(NO_3)_2{\cdot}2{,}5H_2O$ were weighed on an analytical balance, transferred to a 250 ml heat-resistant beaker, and deionized water acidified with nitric acid was added until the salts were completely dissolved. To accelerate the process, a magnet was immersed in the reaction vessel for stirring and the beaker was placed on a magnetic stirrer. The stirring speed was approximately 500 rot/minute.

Then 1.26 g of citric acid monohydrate was added to the resulting transparent solution of zirconyl and hafnyl nitrates (the total molar ratio of metal salts refers to the number of moles of citric acid as 1:1). After that, the resulting solution was placed on a hot plate. After 2 hours of removal of residual moisture, the solution turned into a viscous gel, which, upon further heating on a hotplate at 200°C, burned to form a dry residue (**Fig. A1**) due to a self-initiated redox reaction between nitrate groups (oxidant) and citric acid molecules (reducer). Then, the flakes after burning the gel (**Fig. A2**) were ground in an agate mortar. The ground powder of pale-yellow color (**Fig. A3**) was placed in alundum crucibles and calcined at different temperatures: 500°C, 600°C, 700°C and 800°C.

| 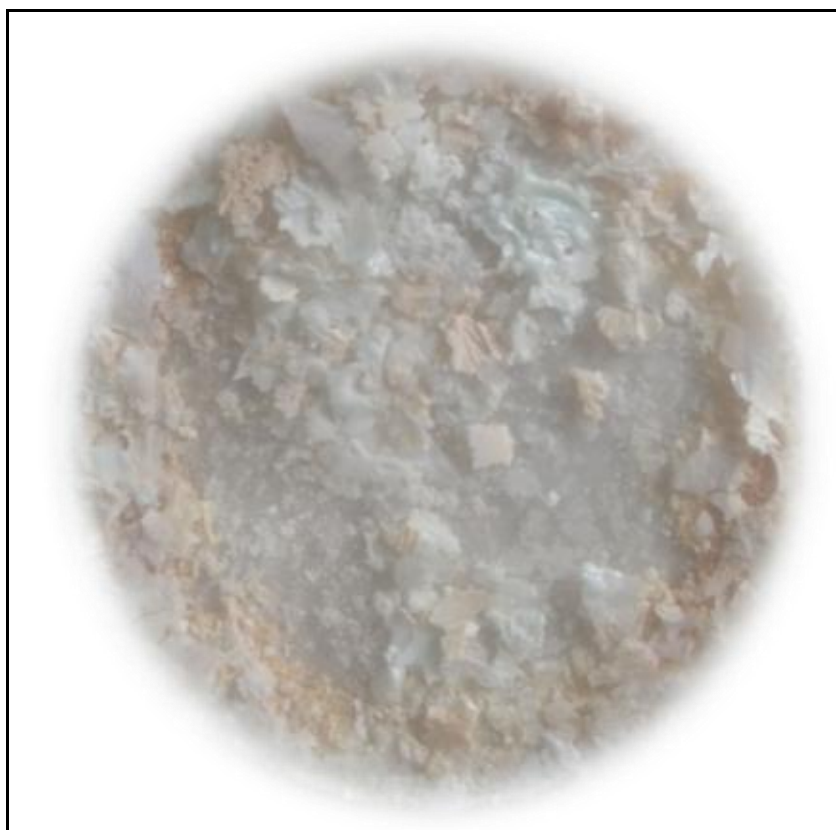 | 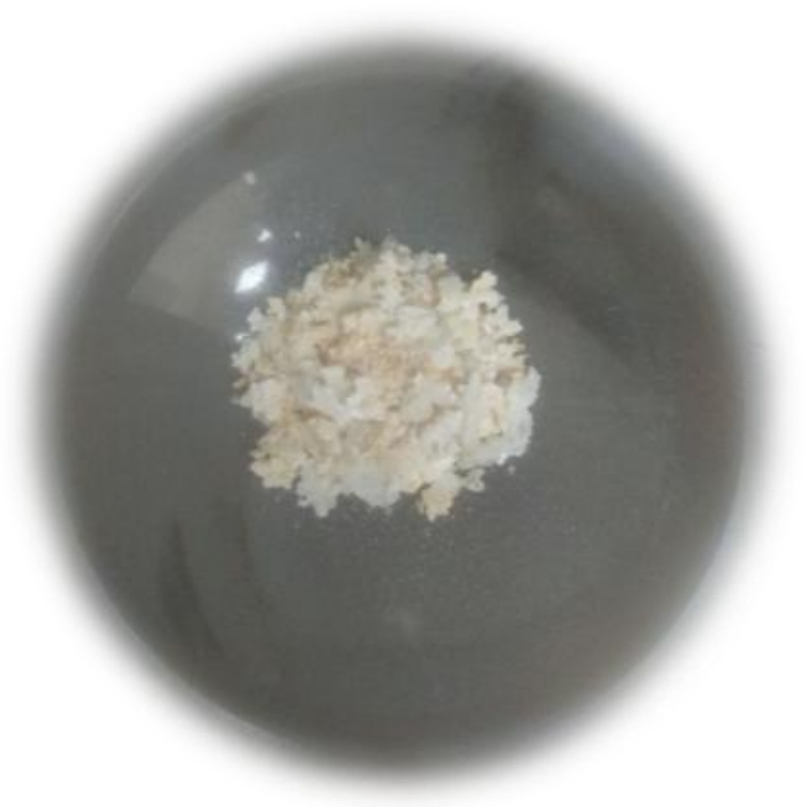 | 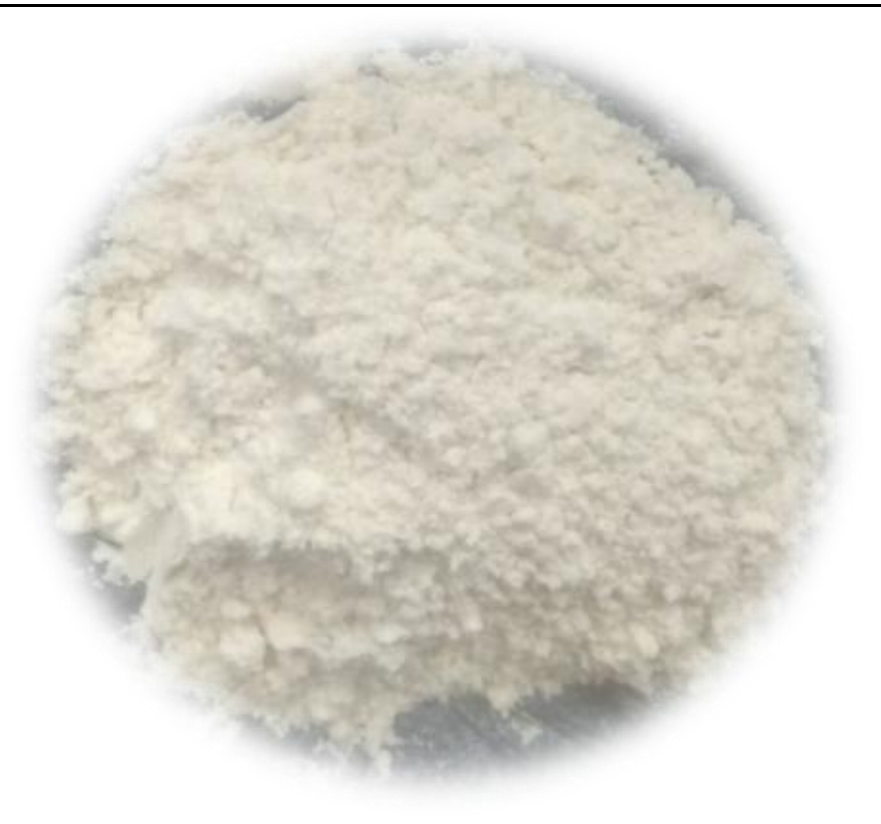 |
| --- | --- | --- |
| **Fig. A1.** The solid residue after burning the dried gel at the bottom of the beaker. | **Fig. A2.** Flakes formed after burning the gel in an agate mortar. | **Fig. A3.** Dispersed powder after burning the gel. |

Sample # 1: powder after gel combustion, calcined at 500°C for 200 min. in a muffle furnace SNOL 8.2/1100. Heating rate – 5°C/min. Cooling is spontaneous, together with the furnace. Powder color – black.

Sample # 2: powder after gel combustion, calcined at 600°C for 200 min. in a muffle furnace SNOL 8.2/1100. Heating rate – 5°C/min. Cooling is spontaneous, together with the furnace. Powder color – off-white.

Sample # 3: powder after gel combustion, calcined at 700°C for 200 min. in a muffle furnace SNOL 8.2/1100. Heating rate – 5°C/min. Cooling is spontaneous, together with the furnace. Powder color – white.

Sample # 4: powder after gel combustion, calcined at 800°C for 200 min. in a muffle furnace SNOL 8.2/1100. Heating rate – 5°C/min. Cooling is spontaneous, together with the furnace. Powder color – white.

## APPENDIX B. Differential thermal analysis, thermogravimetry (DTA/TG) and X-ray diffraction studies

### B1. Differential thermal analysis and thermogravimetry (DTA/TG)

To control the phase composition and study the processes occurring during the synthesis of substituted hafnium (IV) oxide, the method of synchronous differential thermal analysis and thermogravimetry (DTA/TG) in combination with X-ray diffraction studies was used. Thermogravimetric (TG) and differential thermal analysis (DTA) of the powder after gel combustion were performed on an automatic synchronous thermal analyzer DTG-60H (Shimadzu, Japan).

TG/DTA curves were recorded in the range of 25 – 800 °C with a heating rate of 5°C/min. A sample weighing 5–10 mg was placed in an alundum crucible. The experiments were performed in an air atmosphere at a flow rate of 100 ml/min. Pure aluminum oxide (α-alumina, α-$Al_2O_3$) was used as a reference sample. On the derivatogram (**Fig. B1**) of the dry residue after combustion of nitrate-citrate gel, three regions can be conditionally distinguished, which are accompanied by mass loss on the TG curve: from 20 °C to 120 °C (region I), from 120 °C to 450 °C (region II) and from 450 °C to 600 °C (region III).

As can be seen from **Fig. B1**, the first stage of thermal decomposition is associated with a slight broad endo-effect on the DTA curve at 63 - 100 °C. At the same time, adsorbed and chemically bound water molecules are removed. The mass loss at stage I is 9.4 %. Further increase in temperature at stage II is observed burning out of trace amounts of the organic phase in the presence of nitrate group residues and oxygen of the working atmosphere. In this case, the process of thermo-oxidative decomposition of the sample is accompanied by a very strong exo-effect with a maximum at 412 °C. From the derivatograms, it is clearly visible that the DTA peak at 350-450 °C is a superposition of two processes that occur with the release of heat and are associated with the formation of gaseous products CO, $CO_2$, NO, $N_2$, etc. The mass loss at stage II is 40.7%. At the final stage III, thermochemical transformations continue in the presence of atmospheric oxygen. The exo-thermal effect at 620 °C is most likely due to the oxidation of carbon residues. Therefore, it can be assumed that the black color of the sample #1, synthesized at a much lower temperature of 500°C, is evidence of the presence of carbon in the composition. This assumption is also further confirmed by the data of X-ray powder diffraction. Above 620 °C, the TG curve reaches a plateau, indicating the formation of an oxide matrix.

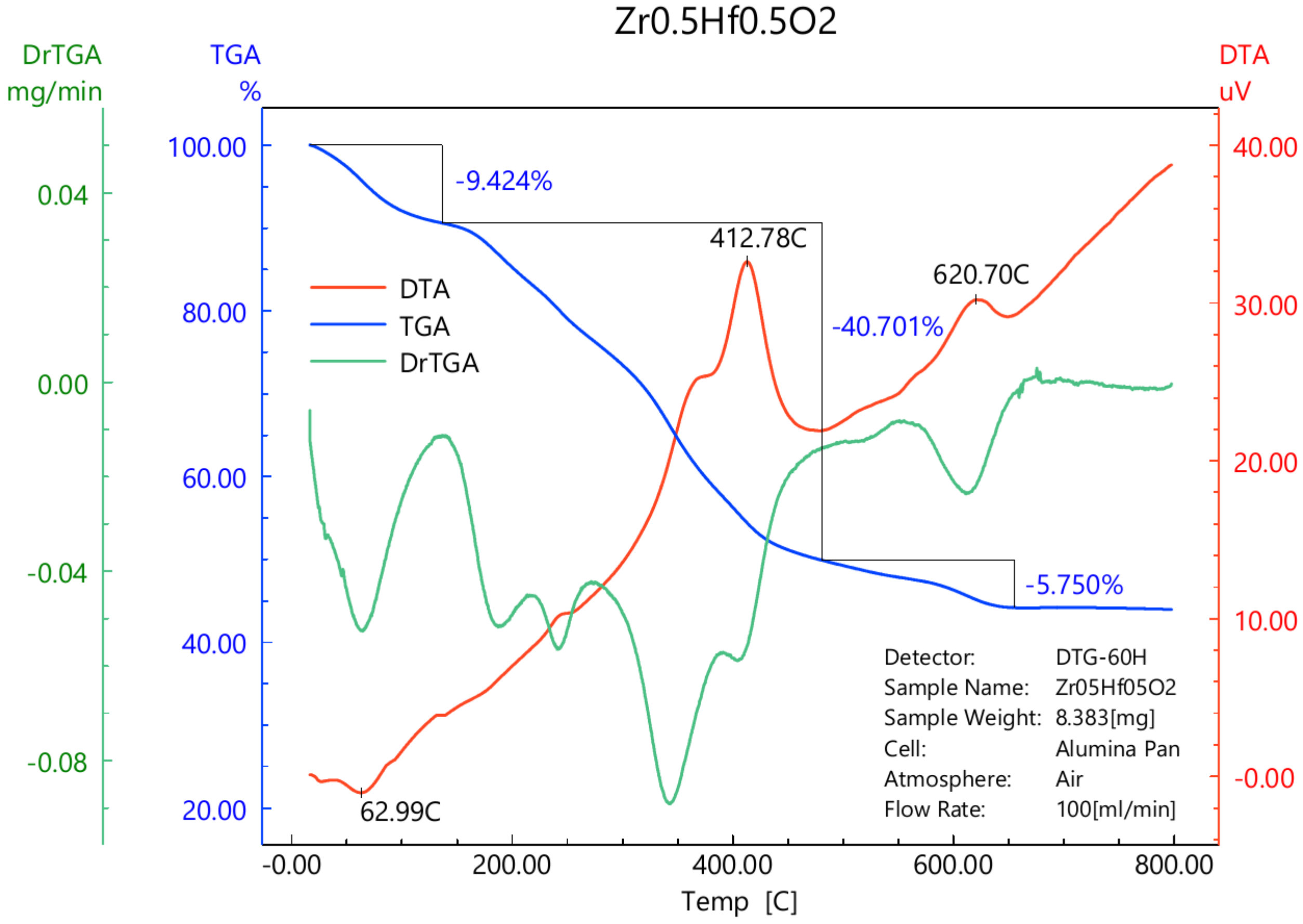

**Figure B1.** The TG/DTA curve of the powder after the gel combustion.

## B2. X-ray diffraction spectra of the HZO nanoparticles

X-ray studies were performed at room temperature on a LabX XRD-6000 diffractometer (Shimadzu, Japan) using CuKα radiation (λ = 1.5406 Å). The diffraction spectrum for phase analysis was recorded using the $\theta - 2\theta$-scanning scheme with Bragg-Brentano focusing in the 2θ angle range from 5 to 70° with a step of Δ(2θ) = 0.02°. The exposure time at each point was 1 second. Measurements were performed at a tube voltage of 35 kV and a current of 30 mA. The phase composition of the $Hf_xZr_{1-x}O_2$ nanopowder for x=1, 0.5 and 0 was identified using the Match! program (version 1.8a) with a built-in database of powder standards Powder Diffraction File (PDF-2). X-ray diffraction patterns were processed by the full-profile Rietveld method using the Match3 computer program with the built-in open crystallographic database COD (Crystallography Open Database – http://www.crystallography.net/cod/ – 512,072 cards) and the FullProf Suite Program (version July 2017). The initial expansion basis contains the monoclinic phase (space group P21/c), the tetragonal phase (space group $P4_2/nmc$) and the orthorhombic phases (space groups Pbca, Pbcm and Pca21). As determined from XRD analysis, the $Hf_{0.5}Zr_{0.5}O_2$ samples # 1 – 4 contain the monoclinic phase and/or orthorhombic phase, the $ZrO_2$ sample # 5 contains the monoclinic phase and about 16 vol.% of the tetragonal phase (space group $P4_2/nmc$), the $HfO_2$ sample # 6 contains the 98 % of monoclinic phase and about 2 vol.% of the orthorhombic phase (see **Table B1** and **Table 1** in the main text). The value of the theoretical density $\rho_{XRD}$ of the samples was calculated from the powder X-ray diffraction data using the formula $\rho_{XRD} = \frac{Z \cdot M_r}{V \cdot N_A}$, where Z is the number of formula units of the unit cell (for the o-phase Z = 8, and for the m-phase Z = 4); Mr is the molecular weight of zirconium-substituted hafnium oxide $Hf_{0.5}Zr_{0.5}O_2$ (g/mol); $V$ is the volume of the unit cell (in $cm^3$); $N_A$ is Avogadro's number ($6.022 \cdot 10^{23}$ $mol^{-1}$). The Rietveld-corrected diffraction patterns of zirconium-hafnium oxide powders obtained at room temperature are shown in **Fig. 1** (main text).

According to the results of phase analysis of the samples, it was established that the process of formation of the crystalline phase begins at 500°C. However, the hafnium-zirconium oxide still contains an X-ray amorphous carbon-containing component, the presence of which worsens the determination of crystallographic parameters and sizes of scattering crystallites. Thus, the temperature of 500°C is insufficient for burning out the organic phase residues in the dry nitrate-citrate gel. In general, the diffraction patterns of all the studied systems contain diffraction reflections belonging to the orthorhombic phases $Hf_{0.5}Zr_{0.5}O_2$ and monoclinic $Hf_{0.5}Zr_{0.5}O_2$. The volume fractions of these phases significantly depend on the calcination temperature: at 500 °C and 600 °C, the content of the orthorhombic phase varies within 90%, and at 800 °C it does not exceed 7%. It is worth noting that it was not possible to obtain 100% pure $Hf_{0.5}Zr_{0.5}O_2$ with an orthorhombic crystal lattice. The crystallographic parameters, refined by the full-profile Rietveld method, are given in **Table B1**.

**Table B1.** Phase composition of HZO nanoparticles

| **Sample** | **HZO-500** | | **HZO-600** | | **HZO-700** | | **HZO-800** | |
|---|---|---|---|---|---|---|---|---|
| Sintering temperature, T (°C) | 500 | | 600 | | 700 | | 800 | |
| Phase composition | m-$Hf_{0.5}Zr_{0.5}O_2$ (15±4%) | o-$Hf_{0.5}Zr_{0.5}O_2$ (85±11%) | m-$Hf_{0.5}Zr_{0.5}O_2$ (9±1%) | o-$Hf_{0.5}Zr_{0.5}O_2$ (91±4%) | m-$Hf_{0.5}Zr_{0.5}O_2$ (58±3%) | o-$Hf_{0.5}Zr_{0.5}O_2$ (42±5%) | m-$Hf_{0.5}Zr_{0.5}O_2$ (93±2%) | o-$Hf_{0.5}Zr_{0.5}O_2$ (7±1%) |
| Crystal system | mono-clinic | ortho-rhombic | mono-clinic | ortho-rhombic | mono-clinic | ortho-rhombic | mono-clinic | ortho-rhombic |
| Space group | $P2_1/c$ | $Pbca+Pbcm+Pca2_1$ | $P2_1/c$ | $Pbca+Pbcm+Pca2_1$ | $P2_1/c$ | $Pbca+Pbcm+Pca2_1$ | $P2_1/c$ | $Pbca+Pbcm+Pca2_1$ |
| Lattice parameters $a, b, c$, (Å) | $a$ = 5.091(6) Å; $b$ = 5.184(8) Å; $c$ = 5.317(9) Å; | $a$ = 10.184(18) Å; $b$ = 5.173(3) Å; $c$ = 5.099(8) Å; | $a$ = 5.263(3) Å; $b$ = 5.289(5) Å; $c$ = 5.401(4) Å; | $a$ = 10.273(12) Å; $b$ = 5.272(2) Å; $c$ = 5.163(6) Å; | $a$ = 5.132(1) Å; $b$ = 5.165(1) Å; $c$ = 5.310(1) Å; | $a$ = 10.224(15) Å; $b$ = 5.167(5) Å; $c$ = 5.076(4) Å; | $a$ = 5.135(1) Å; $b$ = 5.177(1) Å; $c$ = 5.312(1) Å; | $a$ = 10.096(12) Å; $b$ = 5.193(6) Å; $c$ = 5.110(6) Å; |
| Lattice angle $\alpha, \beta, \gamma$ (°) | $\alpha = \gamma =$ 90°; $\beta =$ 99.15(9)° | $\alpha = \beta = \gamma =$ 90° | $\alpha = \gamma =$ 90°; $\beta =$ 99.18(3)° | $\alpha = \beta = \gamma =$ 90° | $\alpha = \gamma =$ 90°; $\beta =$ 99.15(1)° | $\alpha = \beta = \gamma =$ 90° | $\alpha = \gamma =$ 90°; $\beta =$ 99.14(1)° | $\alpha = \beta = \gamma =$ 90° |
| Unit cell volume, $V$ | 138.57(38) | 268.69(64) | 148.44(22) | 279.68(51) | 138.97(5) | 268.20(52) | 139.44(3) | 267.95(54) |

| (Å$^3$) | | | | | | | | |
|---|---|---|---|---|---|---|---|---|
| Number of formula units per cell, *Z* | 4 | 8 | 4 | 8 | 4 | 8 | 4 | 8 |
| X-Ray density, $\rho_{XRD}$ (g/cm$^3$) | 7.998 | 8.249 | 7.566 | 7.925 | 7.975 | 8.265 | 7.948 | 8.272 |
| $R_p$ (%) | 8.6 | | 9.8 | | 8.6 | | 8.2 | |
| $R_{wp}$ (%) | 10.9 | | 12.6 | | 11.3 | | 10.9 | |
| $R_{exp}$ (%) | 10.1 | | 9.5 | | 9.5 | | 9.9 | |
| $\chi^2$ (%) | 1.18 | | 1.75 | | 1.43 | | 1.20 | |

## APPENDIX C. Details of electrophysical measurements

To study the electronic transport in $Hf_{0.5}Zr_{0.5}O_2$ nanopowders the powders were placed in a Teflon cell between two brass plungers that determined the size of the sample, served to create uniaxial pressure, and also served as electrical contacts. The diameter of the samples was 4 mm, the distance between the contacts was 100±10 μm. The sample was placed on a stand for measuring electrical characteristics. The voltage sweep on the sample was carried out by a software-controlled Instek PSP 603 power supply with a step of 20 mV every 2 s, the registration of the voltage drop across the sample and the reference resistor was carried out using a precision multimeter Keithley 2000-SCAN, every 2 s with recording on a PC. The sweep time was selected in such a way as to reduce the influence of transient (dynamic) processes on the measurement of the I-V curves, the total measurement time for one sample in the DC mode was more than 4000 s. To measure the capacitance of the studied cells in the range ($4 – 5\cdot10^5$) Hz, an RLC-meter LCX200 ROHDE & SCHWARZ was used. The measurements were carried out at 290 K. The numbers of the samples # 1 – 6 are given in **Table 1** (main text).

**Figure C1** shows the frequency dependences of the capacitance $C_p$ and the dielectric loss tangent of the samples #1-6. The sample # 1 demonstrates a weak dependence of the capacitance on the frequency and the level of the applied ac voltage, the loss tangent angle varies from 0.5 to 0.1, which may be additional evidence that it is amorphous. A strong dependence of the capacitance on the frequency is characteristic for the samples # 2-6. In particular, the capacitance value decreases with frequency by more than 100 times for the sample # 6, and the capacitance decreases in several times with an increase in voltage from 0.5 to 5 V for 100 Hz. Also, a non-monotonic dependence of the dielectric loss tangent on the frequency is observed for these samples.

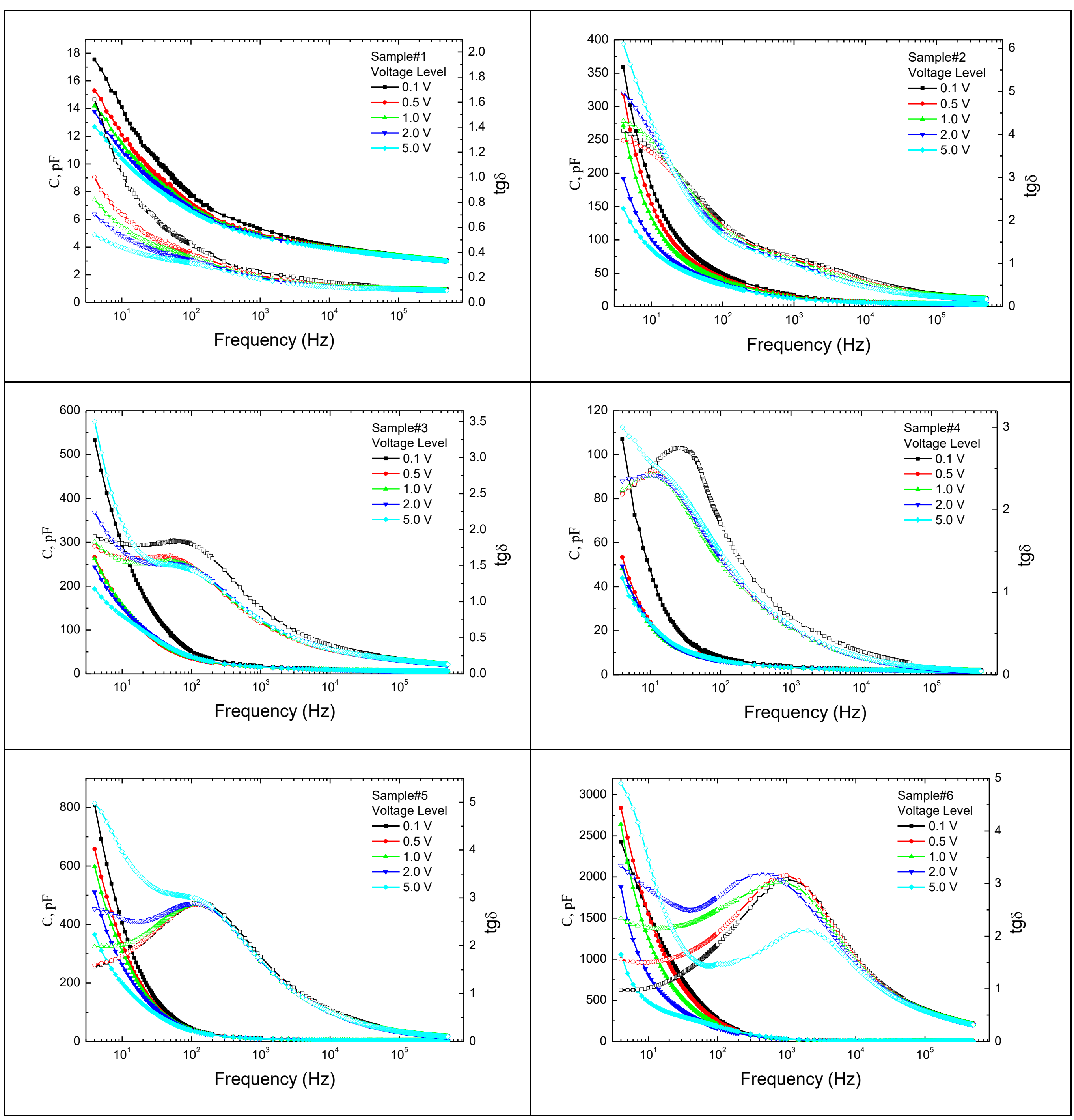

**Figure C1.** Frequency dependences of capacitance $C_p$ (filled symbols) and dielectric loss tangent (empty symbols) for the $Hf_{0.5}Zr_{0.5}O_2$ nanopowders (samples #1 – 4), $ZrO_2$ nanopowder (sample # 5) and $HfO_2$ nanopowder (sample #6).

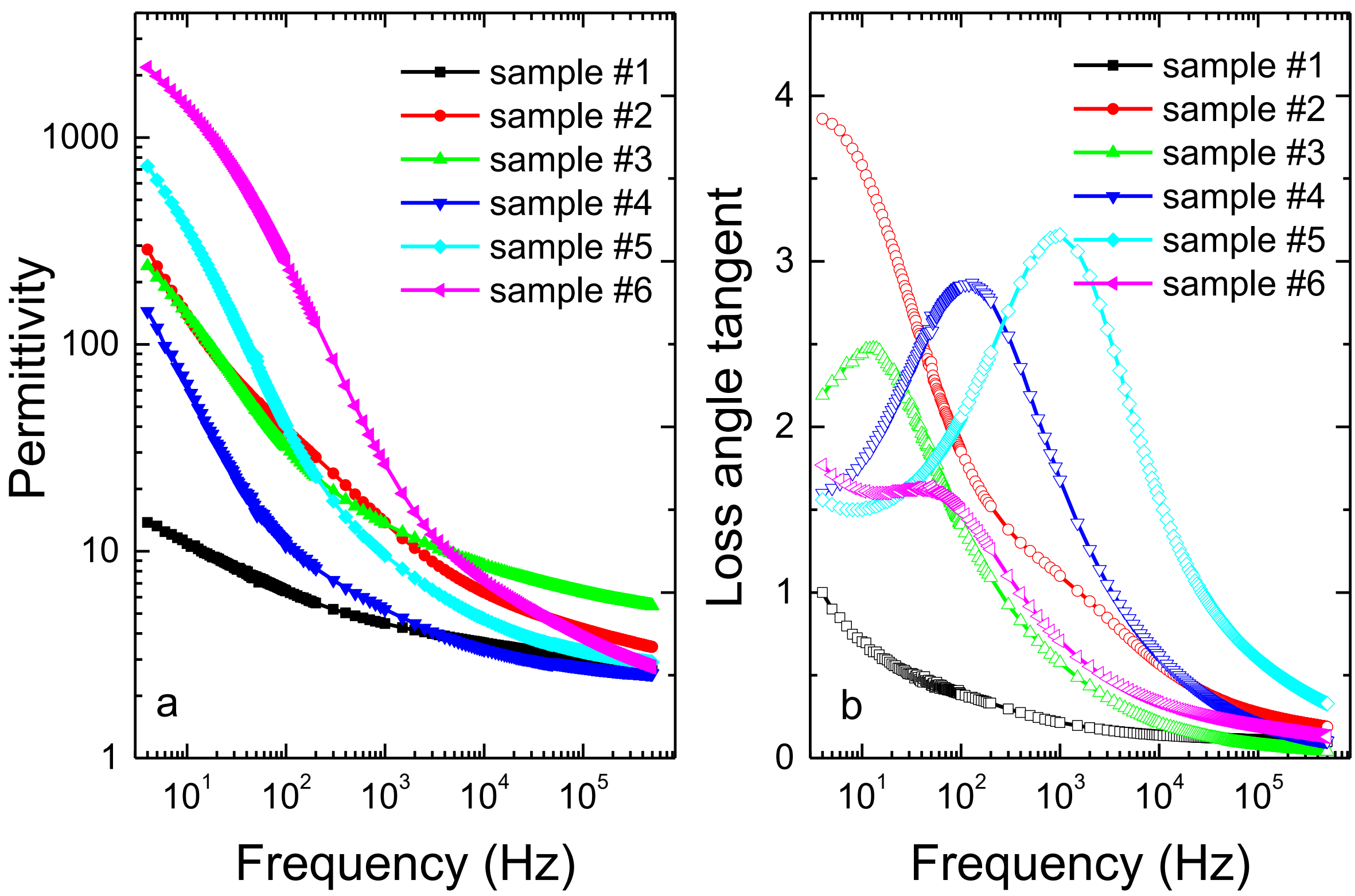

**Figure C2.** Frequency dependences of effective dielectric permittivity ε **(a)** and the loss tangent angle **(b)** for the $Hf_{0.5}Zr_{0.5}O_2$ nanopowders (samples #1 – 4), $ZrO_2$ nanopowder (sample # 5) and $HfO_2$ nanopowder (sample #6). The ac voltage amplitude is 0.5 V.

## APPENDIX D. Details of the density functional theory calculations

The density functional theory (DFT) calculations were performed using the full-potential all-electron local-orbital (FPLO) code [63] for the local density approximation (LDA) functional. We have used the default FPLO basis. The structural parameters of 12-atom cells are optimized with using local density approximation (LDA), a $6\times6\times6$ *k*-point mesh for Brillouin zone sampling. An energy convergence threshold is $10^{-8}$ Ha, and a force convergence threshold is $10^{-3}$ eV/Å. Both lattice constants and ionic positions are fully relaxed.

The calculated equilibrium lattice parameters *a*, *b*, and *c* were compared with experimental results [64, 65, 66] and values determined by reference calculation packages [67] (see **Table D1**). It appeared that the crystal structure with the monoclinic symmetry has the lowest ground state energy in the bulk $Hf_{0.5}Zr_{0.5}O_2$ (HZO), as well as in $HfO_2$ and $ZrO_2$. At the same time, the differences between the energies of the ferroelectric orthorhombic phases (f-phase) and the monoclinic phase (m-phase), calculated by the non pseudopotential all-electron FPLO package, turned out to be smaller than those calculated by other DFT methods. It is also seen from **Table D1** that the difference between the m-HZO and f-HZO (~7.8 meV/f.u.) is significantly smaller than the difference between the m-$HfO_2$ and f-$HfO_2$ (~11.6 meV/f.u.). Thus, the addition of Zr may stabilize the f-phase.

**Table D1.** Comparison of the lattice constants *a*, *b* and *c* (in Å) of a 12 atomic cell and total energy difference $\Delta U$ between the m- and f- phase cells (relative to the m-phase) in meV/f.u. for HZO, $HfO_2$ and $ZrO_2$ compounds

| Structure | $\Delta U$ (meV/f.u.) | $V$(Å$^3$) | *a* (Å) | *b* (Å) | *c* (Å) | Method [reference] |
|---|---|---|---|---|---|---|
| m-HZO | 0 | 135.9 | 5.09 | 5.16 | 5.25 | LDA[a] |
| | 0 | 137.6 | 5.11 | 5.18 | 5.28 | LDA [67] |
| f-HZO | 7.8 | 131.3 | 5.01 | 5.21 | 5.03 | LDA[a] |
| | 49 | 132.8 | 5.03 | 5.23 | 5.05 | LDA [67] |
| | - | 132.3 | 5.01 | 5.24 | 5.04 | Exp. [64] |
| m-$HfO_2$ | 0 | 134.6 | 5.07 | 5.14 | 5.24 | LDA[a] |
| | 0 | 137.1 | 5.11 | 5.16 | 5.28 | LDA [67] |
| | - | 135.8 | 5.07 | 5.14 | 5.29 | Exp. [64] |
| f-$HfO_2$ | 11.6 | 129.5 | 4.99 | 5.18 | 5.01 | LDA[a] |
| | 62 | 132.1 | 5.02 | 5.22 | 5.04 | LDA [67] |
| m-$ZrO_2$ | 0 | 138.6 | 5.12 | 5.20 | 5.28 | LDA[a] |
| | 0 | 138.2 | 5.11 | 5.20 | 5.28 | LDA [67] |
| | - | 140.3 | 5.15 | 5.20 | 5.32 | Exp. [65] |
| f-$ZrO_2$ | 6.6 | 133.7 | 5.05 | 5.22 | 5.07 | LDA[a] |
| | 37 | 133.4 | 5.04 | 5.24 | 5.05 | LDA [67] |
| | - | 135.5 | 5.07 | 5.26 | 5.08 | Exp. [66] |

[a] This work